\documentclass[twocolumn,english,pra]{revtex4-2}
\usepackage[T1]{fontenc}
\usepackage[latin9]{inputenc}
\usepackage{xcolor}
\usepackage{float}
\usepackage{amsmath}
\usepackage{graphicx}
\usepackage{url}
\usepackage{hyperref}
\makeatletter

\newcommand{\lyxmathsym}[1]{\ifmmode\begingroup\def\b@ld{bold}
  \text{\ifx\math@version\b@ld\bfseries\fi#1}\endgroup\else#1\fi}

\usepackage{hyperref}

\makeatother

\usepackage{babel}
\begin{document}
\title{Interaction-induced reentrance of Bose glass and quench dynamics of
Bose gases in twisted bilayer and quasicrystal optical lattices}
\author{Shi-Hao Ding$^{1,2,3}$}
\thanks{These authors contributed equally to this work.}
\author{Li-Jun Lang$^{1,2,3}$}
\thanks{These authors contributed equally to this work.}
\author{Qizhong Zhu$^{2,3}$}
\email{qzzhu@m.scnu.edu.cn}

\author{Liang He$^{1,2,3}$}
\email{liang.he@scnu.edu.cn}

\affiliation{$^{1}$Institute for Theoretical Physics, School of Physics, South
China Normal University, Guangzhou 510006, China}
\affiliation{$^{2}$Key Laboratory of Atomic and Subatomic Structure and Quantum
Control (Ministry of Education), Guangdong Basic Research Center of
Excellence for Structure and Fundamental Interactions of Matter, School
of Physics, South China Normal University, Guangzhou 510006, China}
\affiliation{$^{3}$Guangdong Provincial Key Laboratory of Quantum Engineering
and Quantum Materials, Guangdong-Hong Kong Joint Laboratory of Quantum
Matter, South China Normal University, Guangzhou 510006, China}
\begin{abstract}
We investigate the ground state and dynamical properties of ultracold
gases in optical lattices with a quasicrystal structure---a scenario
inspired by recent experiments on twisted bilayer optical lattices
and optical quasicrystals. Our study reveals that the interplay between
on-site repulsive interactions and a quasiperiodic potential gives
rise to rich physics. At low filling factors, increasing the interaction
strength induces a delocalization effect that transforms a Bose-glass (BG)
phase, characterized by disconnected superfluid (SF) regions, into
a robust SF phase with a percolated network of SF clusters. This transition
is quantitatively characterized by monitoring the percolation probability.
At higher filling factors, we uncover a striking reentrant behavior:
As the on-site interaction increases, the system initially transitions
from BG to SF, but a further increase reverses this trend, returning
the system to the BG phase. This reentrance is ascribed to an interaction-driven
rearrangement of particles, where a once percolated SF network fragments
into isolated SF islands as repulsive interactions dominate. Furthermore,
our analysis of quench dynamics demonstrates distinct transient behaviors.
Intra-phase quenches yield minimal variations in both the percolation
probability and the inverse participation ratio (IPR) of the particle
density distribution. In contrast, interphase quenches produce pronounced
effects; for instance, a quench from the SF to BG phase is marked
by an abrupt loss of global SF connectivity, while a BG-to-SF quench
features oscillatory changes in the percolation probability and a
gradual decrease in the IPR, eventually stabilizing the SF phase.
Our findings unveil the complex interplay between quasiperiodic optical
lattice potential and interaction in ultracold Bose gases, offering
valuable insights that are highly pertinent to current experimental
efforts employing state-of-the-art twisted bilayer and quasicrystalline
optical lattice platforms.
\end{abstract}
\maketitle

\section{Introduction}

Recently, twisted bilayer materials have opened a brand-new era of twistronics,
which has become fertile ground for studying and engineering strong-correlation phenomena \citep{cao_Nature_2018_Correlated_Insulator,li_Nature_2021_Continuous_Mott,cao_Nature_2018_Unconventional_SC,gong_Nature_2017,huang_Nature_2017,chen_Nature_2020,li_Hongyuan_Nature_2021,regan_Nature_2020,serlin_Science_2020,li_Nature_2021_QAH,zhou_Nature_2021a}.
Most research has focused on moir\'e patterns, the commensurate patterns
of multiple twisted crystal layers, in which the whole system retains
certain translational symmetry but the unit cell is much larger than
the original one, leading to the flattening of bands and thus enhancing
interaction effects. Experimentalists have observed in the laboratory
many emergent strong-correlation effects, such as correlated insulators
\citep{cao_Nature_2018_Correlated_Insulator,li_Nature_2021_Continuous_Mott},
superconductivity \citep{cao_Nature_2018_Unconventional_SC}, magnetism
\citep{gong_Nature_2017,huang_Nature_2017,chen_Nature_2020}, generalized
Wigner crystals \citep{li_Hongyuan_Nature_2021,regan_Nature_2020},
and integer and fractional quantum anomalous Hall insulators \citep{serlin_Science_2020,li_Nature_2021_QAH,zhou_Nature_2021a,Cai_Nature_2023_add_1,Park_Nature_2023_add_2,Zeng_2023_add_3,Xu_2023_PRX_add_4,Lu_2024_add_5}.

On the other hand, quasicrystal patterns can emerge when the crystal
layers are twisted in an incommensurate way that completely breaks
the translational symmetry but preserves the long-range order \citep{Shechtman_Cahn_1984_PRL},
thereby mediating between order and disorder. Quasicrystals
have been experimentally realized and highly tuned by twisting three
layers of graphene \citep{Uri_Jarillo-Herrero_2023_Nature} or bilayers
of tungsten diselenides \citep{Li_Shih_2024_Nature} but have received less attention before \citep{Park_Lee_2019_PRB,Huang_Liu_2019_PRB}
than the commensurate case. It is well-known that a quasicrystal will
lead to Anderson localization \citep{Anderson_1958_PR}, a phenomenon
that lacks the diffusiveness of noninteracting particles when it is subjected
to a random or quasiperiodic potential landscape. In the one-dimensional
(1D) and two-dimensional (2D) cases, all single-particle eigenstates
can be localized by a finite strength of the quasicrystal potential \citep{Aubry_Andre_1980_AIPS,Roati_Inguscio_2008_Nature,Lahini_Silberberg_2009_PRL},
different from the random disorder where the localization occurs for
any infinitesimal strength \citep{Abrahams_Ramakrishnan_1979_PRL}.
Therefore, quasicrystals are ideal platforms for studying localization
physics.

Now that the moir\'e pattern can induce many exotic strongly correlated
phenomena, one may wonder how the localization feature in twisted
bilayer quasicrystals competes with the interaction. 
In fermionic systems, most research has focused on the thermal properties of many-body
localization, especially in the 1D case \citep{Iyer_Huse_2013_PRB,Mondaini_Rigol_2015_PRA,Xu_DasSarma_2019_PRR,Vu_DasSarma_2022_PRL,Wang_Yu_2021_PRL}.
The incommensurability is also found to radically change the ground
state of interacting fermions, compared with the commensurate case
\citep{Vu_DasSarma_2021_PRL,Goncalves_Ribeiro_2024_NatPhys,Kraus_Berkovits_2014_PRB,Cookmeyer_Moore_2020_PRB,Oliveira_Amorim_2024_arXiv,Gonccalves_Ribeiro_2024_PRB}.
In bosonic systems, the interplay of the randomness or quasiperiodicity
and the interaction will lead to an exotic many-body ground state---Bose
glass (BG) \citep{Giamarchi_Schulz_1988_PRB,Fisher_Fisher_1989_PRB},
which is insulating like an Anderson insulator but compressible like
a superfluid. Disordered interacting bosons in different dimensions
have been studied using various experimental platforms, such as helium
films on porous substrates \citep{Crowell_Reppy_1997_PRB}, disordered
superconducting films \citep{Sacepe_Ioffe_2011_NatPhys}, doped quantum
magnets \citep{Yu_Roscilde_2012_Nature}, and, especially, optical
lattices \citep{Fallani_Inguscio_2007_PRL,Deissler_Inguscio_2010_NatPhys,Pasienski_DeMarco_2010_NatPhys,Gadway_Schneble_2011_PRL,D'Errico_Modugno_2014_PRL,Meldgin_DeMarco_2016_NatPhys}.
There are also plenty of theoretical studies on interacting bosons
in disordered or quasiperiodic potentials \citep{Rapsch_Zwerger_1999_EPL,Lugan_PRL_2007_1,Lugan_PRL_2007_2,Roux_Giamarchi_2008_PRA,Bissbort_Hofstetter_2009_EPL,Soeyler_Svistunov_2011_PRL,Carleo_PRL_2013,Niederle_Rieger_2015_PRA,Zhang_Capogrosso-Sansone_2015_PRA,Gerster_Montangero_2016_NJP,Yao_Sanchez-Palencia_2020_PRL,Johnstone_Duncan_2021_JPA,Gautier_PRL_2021,Zhu_PRL_2023,Zhu_PRA_2024,Molignini_PRR_2025_2}
and even bosons in interaction-induced disordered or quasiperiodic
potentials \citep{Zeng_PRA_2025}. The experimental
realization of twisted bilayer optical lattices \citep{Meng_2023_Nature}
has opened new avenues to investigate interacting bosons in quasicrystals
using concrete experimental platforms \citep{Johnstone_PRR_2024,Johnstone_PRA_2025}.
More recently, 2D octagonal quasicrystalline optical lattices were directly realized in ultracold-atom experiments by using four
light beams in a plane and one beam in the perpendicular direction,
and BG with ultracold atoms in the weakly interacting regime was observed
\citep{Yu_Schneider_2024_Nature}. 

In this work, we study the ground state and dynamical properties of
a class of ultracold bosonic systems realized in optical lattices
with an aperiodic external potential---a setting directly inspired
by recent experiments on twisted bilayer optical lattices \citep{Meng_2023_Nature}
and optical quasicrystals \citep{Yu_Schneider_2024_Nature}. We investigate
how the interplay between on-site repulsive interactions and a quasiperiodic
potential drives the system between a homogeneous superfluid (SF)
phase and a BG phase. At low filling factors, our findings reveal
that an increase in the on-site interaction strength induces a delocalization
effect, transforming a BG phase---with disconnected SF regions lacking
global connectivity---into a robust SF phase characterized by a percolated
network of SF clusters. This phase transition is quantitatively characterized
by monitoring the percolation probability (see Fig.~\ref{fig:low filling}).
At higher filling factors, our results uncover a striking reentrant
behavior (see Fig.~\ref{fig:high filling}): The system initially
undergoes a transition from a BG to an SF phase as the on-site interaction
strength increases, but a further increase in the interaction leads
to a reversal back to the BG phase. This reentrance is attributed
to the interaction-driven rearrangement of particles, wherein a once-percolated SF network eventually fragments into isolated SF islands
as repulsive interactions become dominant. Moreover, our examination
of the system's quench dynamics reveals distinct transient
behaviors (see Fig.~\ref{fig:quench_dynamics}): While intraphase
quenches produce minimal changes in both the percolation probability
and inverse participation ratio (IPR) of the particle-density distribution,
interphase quenches lead to pronounced dynamical effects. Specifically,
a quench from the SF phase to the BG phase results in an abrupt loss of global
SF connectivity, whereas a BG-to-SF quench is marked by oscillatory
changes in the percolation probability and a gradual decrease in the
IPR, ultimately stabilizing the SF phase. These findings unveil the
rich physics of ultracold Bose gases in optical lattices with an aperiodic
external potential, offering valuable insights that are highly relevant
for state-of-the-art experiments employing twisted bilayer or quasicrystalline
optical lattice platforms.

\section{System and Model}

Inspired by the recent experimental progress in realizing twisted
bilayer optical lattice \citep{Meng_2023_Nature} and optical quasicrystal
\citep{Yu_Schneider_2024_Nature} for ultracold atoms, we consider
a single-component ultracold bosonic system, the Hamiltonian of which
takes the form of a ``single-layer'' Bose-Hubbard model with an
additional external on-site potential, i.e., 
\begin{equation}
\hat{H}=-J\sum_{\left\langle i,j\right\rangle }\hat{b}_{i}^{\dagger}\hat{b}_{j}+\frac{U}{2}\sum_{i}\hat{n}_{i}\left(\hat{n}_{i}-1\right)+\sum_{i}M_{i}\hat{n}_{i},\label{Hamiltonian}
\end{equation}
where $\hat{n}_{i}=\hat{b}_{i}^{\dagger}\hat{b}_{i}$ is the occupation
number operator at site $i$, with $\hat{b}_{i}^{\dagger}~(\hat{b}_{i})$
being the corresponding bosonic creation (annihilation) operator;
$U$ is the strength of the on-site interaction; and $J$ is the hopping
amplitude between nearest-neighbor lattice sites (denoted by $\langle\cdots\rangle$).
Here, we focus on the case where the underlying optical lattice is
a square lattice because it is the most relevant for the current experimental
setups \citep{Meng_2023_Nature,Yu_Schneider_2024_Nature}. The last
term of $\hat{H}$ is the additional on-site external potential with
$M_{i}$ being its strength at site $i$. In the experimental setup
that realizes the twisted bilayer optical lattice \citep{Meng_2023_Nature},
the on-site external potential $M_{i}$ takes into account the influence
of the lattice on the other layer in the large-detuning limit, where
the bilayer system can be effectively described by the single-layer
model \citep{Wan_2024_arXiv,Meng_2023_Nature}. While in the experimental
set-up that realizes the optical quasicrystal \citep{Yu_Schneider_2024_Nature},
$M_{i}$ simply corresponds to the additional optical lattice potential
felt by the atoms. In this regard, the physics of the model Hamiltonian
(\ref{Hamiltonian}) that will be discussed in the following is relevant
for ultracold twisted bilayer optical lattice systems \citep{Meng_2023_Nature}
and optical quasicrystal systems \citep{Yu_Schneider_2024_Nature}
in experiments. 

In the following, we focus on the case where $M_{i}$ assumes the
form 
\begin{align}
M_{i}= & M_{r}\left[\sin^{2}\left(i_{x}\pi\cos\theta+i_{y}\pi\sin\theta\right)\right.\nonumber \\
 & \left.+\sin^{2}\left(i_{y}\pi\cos\theta-i_{x}\pi\sin\theta\right)\right],
\end{align}
where $M_{r}$ is the strength of the external potential, $i_{x}$
($i_{y}$) is the index of the $i\textrm{th}$ site of the square
lattice along the $x$ ($y$) direction, and we choose $\theta=45\lyxmathsym{\textdegree}$.
This corresponds to the case in bilayer systems with a twist angle
of $45\lyxmathsym{\textdegree}$ \citep{Meng_2023_Nature} or
the one in the optical quasicrystal system with the optical lattice
formed by superimposing four independent one-dimensional lattices
in the $x$-$y$ plane at a $45\lyxmathsym{\textdegree}$ angle \citep{Yu_Schneider_2024_Nature}. 

The first two terms of the Hamiltonian (\ref{Hamiltonian}) assume
the form of the conventional Bose-Hubbard model \citep{Fisher_Fisher_1989_PRB};
they favor two homogeneous phases, namely, the homogeneous SF phase
(for large $J/U$) and the Mott insulator phase (for small $J/U$ at integer
filling factors) that respect the discrete translation symmetry of
the underlying lattice. The third term of $\hat{H}$ is an aperiodic
or a quasiperiodic external potential that explicitly breaks the lattice
translation symmetry. Recent investigations \citep{Johnstone_Duncan_2021_JPA}
showed that this type of term can give rise to the BG phase that
usually arises in disordered bosonic systems \citep{Fisher_Fisher_1989_PRB}
due to their aperiodicity. In fact, the interplay between on-site
interaction and the quasiperiodic external potential can give rise
to much richer physics associated with the BG phase. Particularly, at
fixed particle densities that are highly relevant for corresponding
experiments, this interplay can give rise to interaction induced reentrant
transitions for BG phases and rich quench dynamics as we shall show
in the following.

\section{Results}

To study the ground state of the system described by Hamiltonian (\ref{Hamiltonian}),
we used the bosonic Gutzwiller variational method \citep{Krauth_PRB_1992,Jaksch_1998_PRL,Lanata_PRB_2012}.
The variational ground state is assumed to take the site-factorized
form $\left|\textrm{GW}\right\rangle =\left|\phi_{1}\right\rangle \otimes\cdots\otimes\left|\phi_{i}\right\rangle \otimes\cdots\otimes\left|\phi_{N_{\textrm{lat}}}\right\rangle $,
where each $\left|\phi_{i}\right\rangle $ represents the state at
the $i$th lattice site and $N_{\textrm{lat}}$ denotes the total
number of sites in the lattice. The individual single-site states
$\left|\phi_{i}\right\rangle $ are expanded as a superposition of
particle-number eigenstates as $\left|\phi_{i}\right\rangle =\sum_{n=0}^{n_{\textrm{max}}}C_{i.n}\left|n\right\rangle _{i}$,
where $n_{\textrm{max}}$ is the maximum number of particles per site
and $C_{i.n}$ are variational parameters that are to be determined
by minimizing the total energy of the system. In this work, we focus
on the ground-state properties of the system at different filling
factors $\rho\equiv N/N_{\textrm{lat}}$, with $N$ being the total
number of particles in the system. We examine the system at different
filling factors to explore how particle density influences the ground-state properties. For the numerical calculations, we set square-lattice
size $L=51$, the local maximum truncation number $n_{\textrm{max}}=10$,
and we employ open boundary conditions for the numerical results if
not otherwise specified in the text. 

\begin{figure}
\begin{centering}
\includegraphics[scale=0.5]{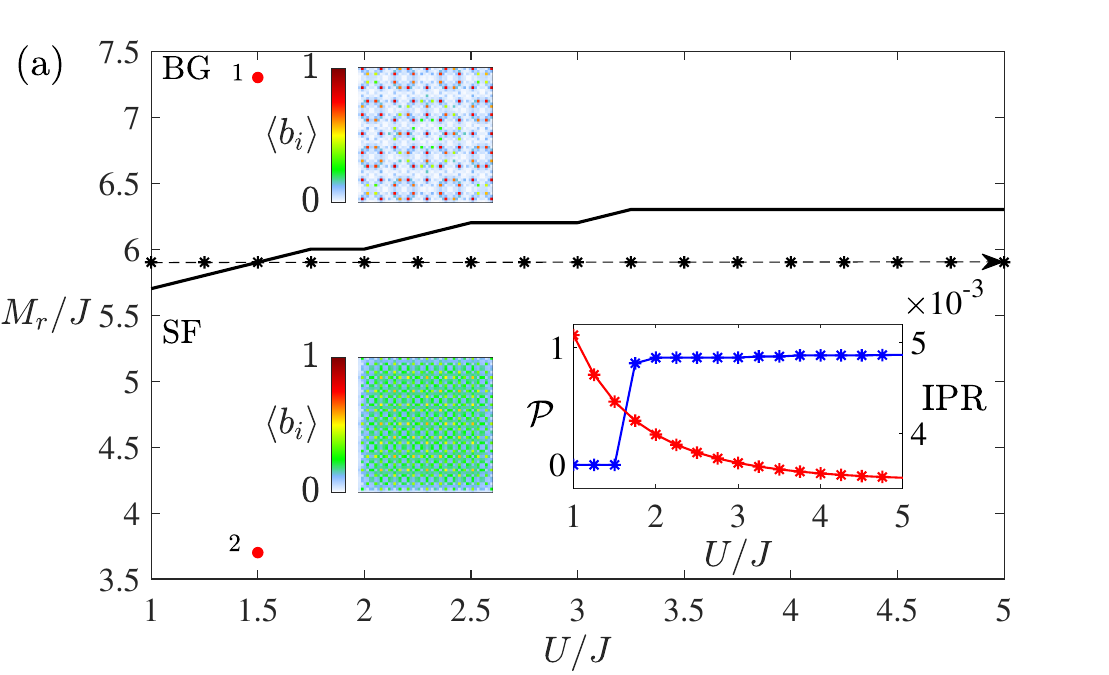}
\par\end{centering}
\begin{centering}
\includegraphics[scale=0.5]{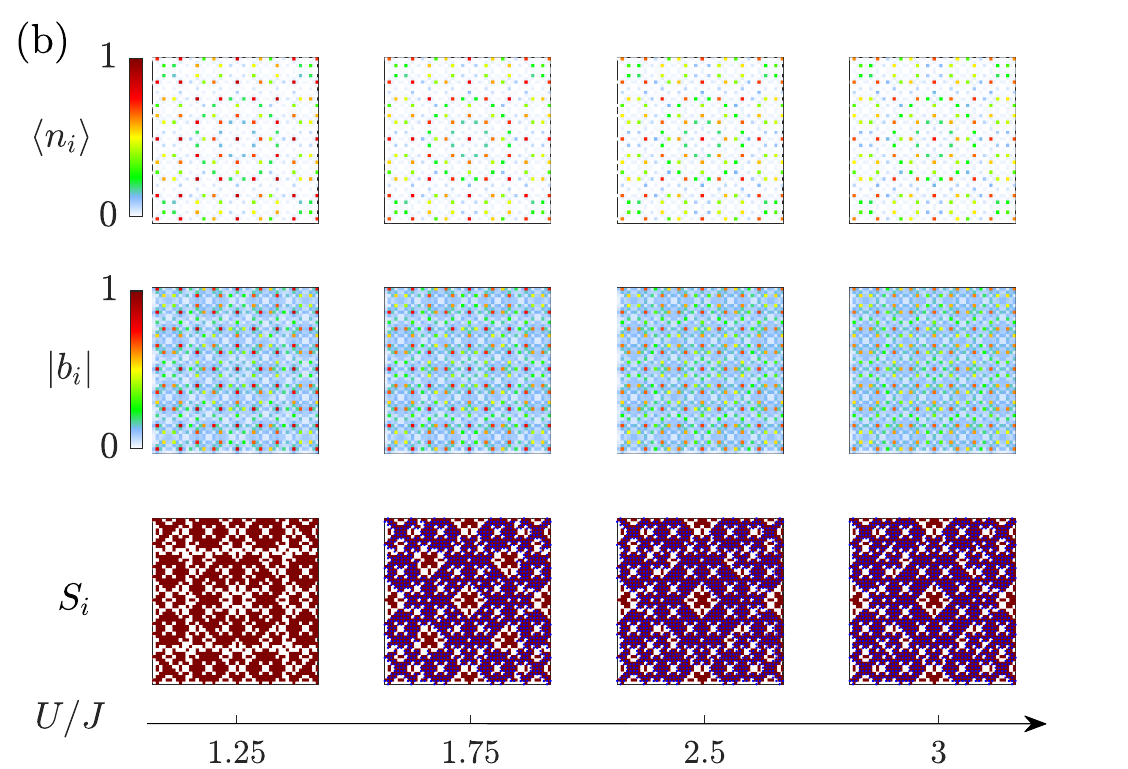}
\par\end{centering}
\caption{\label{fig:low filling}Phase diagram and representative real-space
distributions. (a) Phase diagram for a system of $100$ particles
on a $51\times51$ square lattice, where the black solid line is the
phase boundary between BG (top) and SF (bottom) phases. Inset: Dependence
of the percolation probability $\mathcal{P}$ (blue curve) and the
inverse participation ratio (IPR; red curve) on $U/J$ for points
along the dashed line indicated in the phase diagram. The inserted
SF order-parameter distributions correspond to the nearby red points
marked in the main plot. (b) Real-space distributions of the particle
density $\left\langle \hat{n}_{i}\right\rangle $, the amplitude of
the SF order parameter $\ensuremath{|\langle\hat{b}_{i}\rangle|}$,
and the discrete field $S_{i}$ near the phase boundary ($M_{r}/J=5.9$).
See text for more details. }
\end{figure}

\subsection{Transitions between superfluids and Bose glasses\label{subsec:Transitions-between-superfluids}}

For relatively low filling factors, the typical properties of the system
are summarized in Fig.~\ref{fig:low filling}, which shows the phase
diagram of the system at $\rho\approx0.04$, along with the particle-density distribution $\left\langle \hat{n}_{i}\right\rangle $, the
SF order parameter distribution $\langle\hat{b}_{i}\rangle$, and
the discrete field distribution $S_{i}$. The latter is used to distinguish
between SF and BG phases and will be discussed in detail later. As
shown in the bottom left inset of Fig.~\ref{fig:low filling}(a),
in the low-aperiodic-potential regime, as expected from the homogeneous
Bose-Hubbard model, the system is in the SF phase, with a nonzero
SF order parameter $\langle\hat{b}_{i}\rangle$ that is distributed
relatively evenly across the entire lattice. In contrast, in the high-aperiodic-potential regime, as shown in the top inset of Fig.~\ref{fig:low filling}(a),
the system maintains a nonzero SF order parameter $\langle\hat{b}_{i}\rangle$,
but its distribution exhibits a qualitatively different structure.
In particular, although the SF order parameter is nonzero in various
regions, these regions are \textquotedblleft disconnected\textquotedblright{}
from each other, indicating a transition to a new phase: the BG phase,
which is absent in conventional homogeneous systems of lattice bosons.

To quantitatively distinguish the SF phase from the BG phase, we employ
the percolation method \citep{Sheshadri_1995_PRL,Buonsante_2009_PRA,Kemburi_2012_PRB,Niederle_2013_njp,Johnstone_Duncan_2021_JPA}, a widely used technique for studying phase transitions in disordered
or quasidisordered systems. The key quantity used to differentiate
the SF and BG phases is the percolation probability $\mathcal{P}$,
with $\mathcal{P}=0$ indicating no connected SF region spanning the
entire system (signifying the BG phase) and $\mathcal{P}>0$ indicating
the presence of a percolated SF region (signifying a SF phase spanning
the entire system). In this work, we adopt the method
introduced in \citep{Niederle_2013_njp,Johnstone_Duncan_2021_JPA},
where $\mathcal{P}$ is calculated using the corresponding discrete
field $S_{i}$ \citep{Niederle_2013_njp}, defined as 
\begin{equation}
S_{i}=\begin{cases}
0 & \textrm{if}\textrm{ }I-\gamma_{n}\leq\left\langle \hat{n}_{i}\right\rangle \leq I+\gamma_{n},\\
1 & \textrm{otherwise},
\end{cases}\label{eq:Discrete_field_definition}
\end{equation}
where $\gamma_{n}=5\times10^{-3}$ is the threshold used in the numerical
calculations and $I=0,1,2,\ldots$ denotes non-negative integer values.
This discrete field encapsulates the percolation properties of the
SF regions in the system (see Appendix A for technical details on
extracting $\mathcal{P}$ from $S_{i}$). For example, in the last
row of Fig.~\ref{fig:low filling}(b), red sites represent
various connected SF regions that do not span the entire system, while
blue sites correspond to connected SF regions that do span
the entire system, forming a percolated SF region.

From Fig.~\ref{fig:low filling}(a), we observe that tuning the on-site
interaction strength $U$ can drive the transition between the SF
and BG phases in the high-aperiodic-potential regime. As indicated
by the horizontal arrow in Fig.~\ref{fig:low filling}(a) and the
corresponding distributions for $\left\langle \hat{n}_{i}\right\rangle $,
$\langle\hat{b}_{i}\rangle$, and $S_{i}$ in Fig.~\ref{fig:low filling}(b)
at a fixed aperiodic potential strength $M_{r}$, the system transitions
from the BG phase with $\mathcal{P}=0$ at relatively small $U$ to
the SF phase with a finite percolation probability $\mathcal{P}$
at larger $U$ {[}see the bottom right inset of Fig.~\ref{fig:low filling}(a){]}. 
This transition is driven by a \textquotedblleft delocalization\textquotedblright{}
effect of the repulsive on-site interaction: Under low-filling conditions
($\rho<1$), bosons redistribute from highly occupied sites to nearby
vacant or weakly occupied sites, broadening the density profile. By
contrast, at integer filling, the standard localization effect of
strong interactions can also manifest in this system, driving it into
a Mott insulator (see Fig.~\ref{fig:Avg_SF_vs_U_at_diff_fillings}
and Appendix \ref{appendix_high_filling} for details). As seen in
the bottom right inset of Fig.~\ref{fig:low filling}(a), the IPR
for the normalized density distribution $\widetilde{\left\langle \hat{n}_{i}\right\rangle }\equiv\left\langle \hat{n}_{i}\right\rangle /\sum_{i=1}^{N_{\textrm{lat}}}\left\langle \hat{n}_{i}\right\rangle $
directly characterizes the extent of localization in the system, i.e.,
$\textrm{IPR}\equiv\sum_{i=1}^{N_{\textrm{lat}}}\widetilde{\left\langle \hat{n}_{i}\right\rangle }^{2}$.
The IPR decreases as $U$ increases, reflecting the delocalization
of the particles. At relatively small $U$, the SF regions are localized
{[}see the left most plot for $S_{i}$ in Fig.~\ref{fig:low filling}(b){]}.
As $U$ increases, the SF regions expand, and at large enough $U$,
they can merge, forming a percolated SF region that drives the system
into the SF phase.

\begin{figure}
\begin{centering}
\includegraphics[scale=0.5]{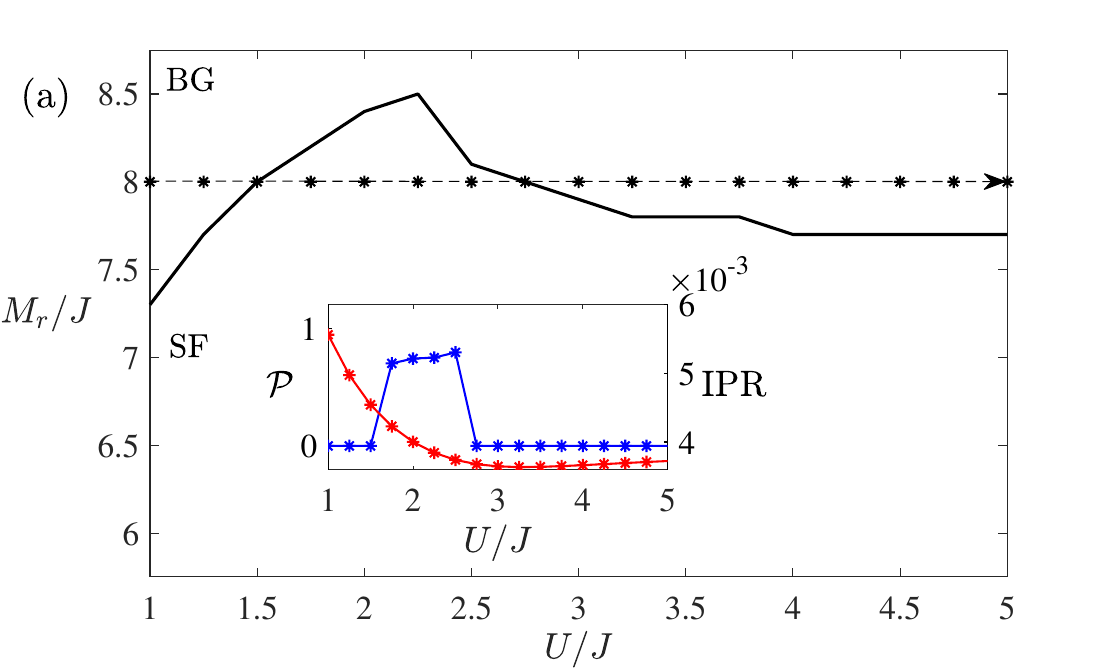}
\par\end{centering}
\begin{centering}
\includegraphics[scale=0.5]{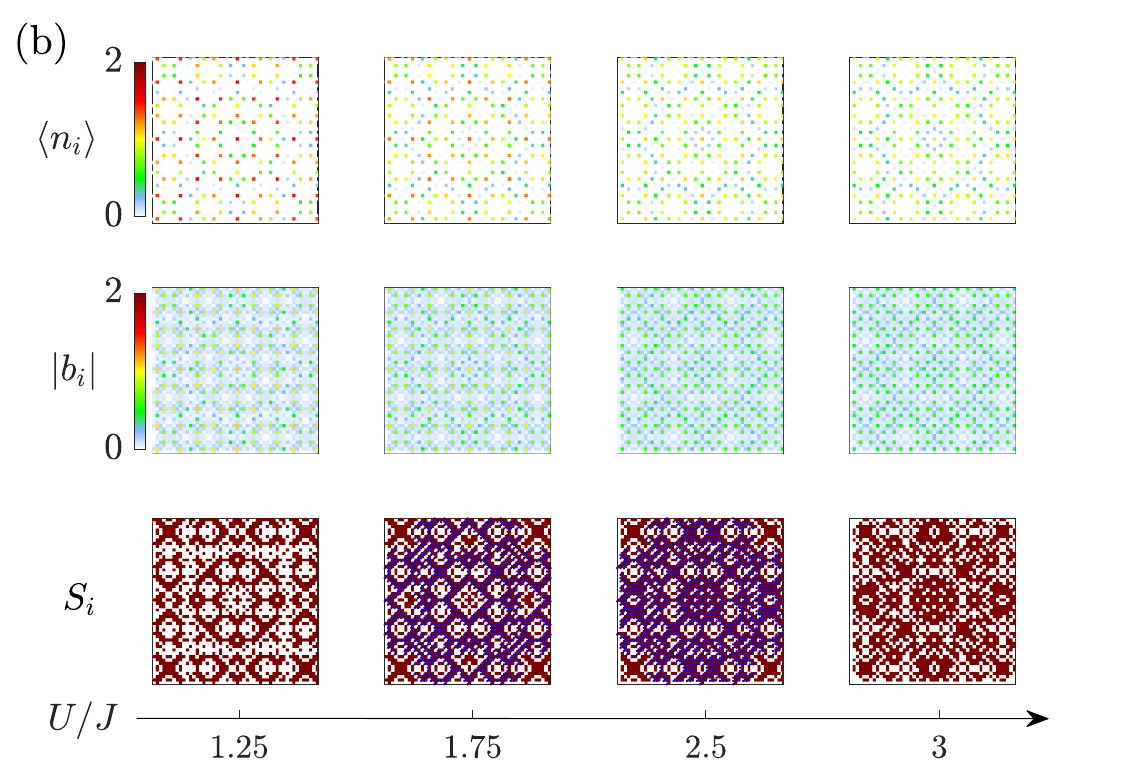}
\par\end{centering}
\caption{\label{fig:high filling}Phase diagram and representative real-space
distributions. (a) Phase diagram for a system of $196$ particles
on a $51\times51$ square lattice, where the black solid line is the
phase boundary between BG (upper) and SF (lower) phases. Inset: Dependence
of the percolation probability $\mathcal{P}$ (blue curve) and IPR
(red curve) on $U/J$ for points along the dashed line indicated in
the phase diagram. (b) Real-space distributions of the particle density
$\left\langle \hat{n}_{i}\right\rangle $, the amplitude of the SF
order parameter $\ensuremath{|\langle\hat{b}_{i}\rangle|}$, and the
discrete field $S_{i}$ near the phase boundary ($M_{r}/J=8$). See
text for more details.}
\end{figure}

\subsection{Reentrance between superfluid and Bose glass phases}

In the previous section, we discussed the system at relatively low
filling factors. From the particle density distribution in Fig.~\ref{fig:low filling}(b),
we observe that the average number of particles per lattice site is
less than 1, indicating the influence of the on-site interaction is
very weak. This thus raises the question of whether different physical
phenomena might arise when the system's filling factor is higher.
To explore this, we present the results for a system with a filling
factor $\rho=0.08$, and the corresponding results are summarized
in Fig.~\ref{fig:high filling}, which includes the phase diagram
and typical real-space distributions. Comparing the results to the phase diagram
at low filling factors, we observe a reentrant transition at higher
filling factors, as shown in Fig.~\ref{fig:high filling}(a). Focusing
on the dashed line in this phase diagram, we see that as the on-site
interaction $U$ increases, the system first transitions from the
BG phase to the SF phase. However, as $U$ continues to increase,
the system transitions back from the SF phase to the BG phase. This
indicates that the on-site interaction can drive a reentrant transition
to the BG phase.

To understand this reentrant behavior, let us first check the typical
real-space distributions of the system as $U$ changes shown in
Fig.~\ref{fig:high filling}(b). When the on-site interaction $U/J=1.25$
is relatively small compared to the external potential field strength
$M_{r}/J=8$, the system consists of isolated SF clusters. As we gradually
increase the on-site interaction to $U/J=1.75$, a percolated SF region
(marked in blue) appears in the $S_{i}$ distribution shown in Fig.~\ref{fig:high filling}(b),
indicating that the system has transitioned from the BG phase to the
SF phase. 

The mechanism of this stage of transition is actually similar to that of the lower-filling case presented in the previous section.
Specifically, when the on-site interaction $U$ increases, the most
significant interaction effect occurs at lattice sites where the average
number of particles is greater than 1. At these sites, the particles
repel each other, leading to a redistribution. As a result, SF paths
emerge during this repulsion process, and the system transitions from
the BG phase to the SF phase.

However, as we further increase $U/J=3$, the system exhibits behavior
different from that at low filling factors. As shown in the bottom
right plot of Fig.~\ref{fig:high filling}(b), the percolated SF
region disappears from the distribution of $S_{i}$ at $U/J=3$, indicating
the system transitions from the SF phase back to the BG phase again.
To understand this stage of transition, we note that although percolated
SF regions can form in the SF phase, the increasing repulsive interaction
strength can still cause the particle density distribution to undergo
further rearrangement, causing percolated SF regions to break up into
isolated SF islands, thus driving the system to enter the BG phase
again. Importantly, this transition is characterized
by the loss of global connectivity of the SF regions, while the local-density distribution changes only slightly. As a result, the IPR remains
small {[}see the inset of Fig.~\ref{fig:high filling}(a){]}, even
though the percolation probability $\mathcal{P}$ drops to zero, signaling
the loss of long-range coherence and the onset of the BG phase. Moreover,
if the particle density of each site is sufficiently low, that would indicate
the local state of each site consists of only a zero- or single-occupation
state. As a consequence, further increasing on-site interaction strength
can hardly impose any physical influence on the system, consistent
with the behavior seen in the last three columns of Fig.~\ref{fig:low filling}(b)
for the low-filling case. In fact, at even high filling factors, one
can observe even more times of reentrance (see Appendix \ref{appendix_high_filling}
for more numerical results at even higher filling factors).\textcolor{red}{{}
}

\begin{figure*}
\centering
\begin{centering}
\includegraphics[totalheight=1.7in]{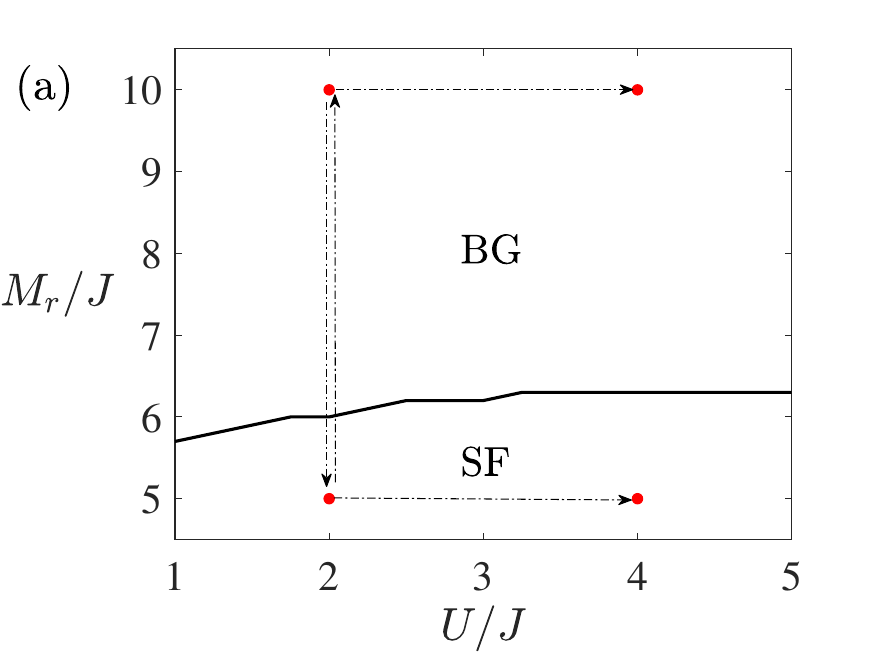}\includegraphics[totalheight=1.7in]{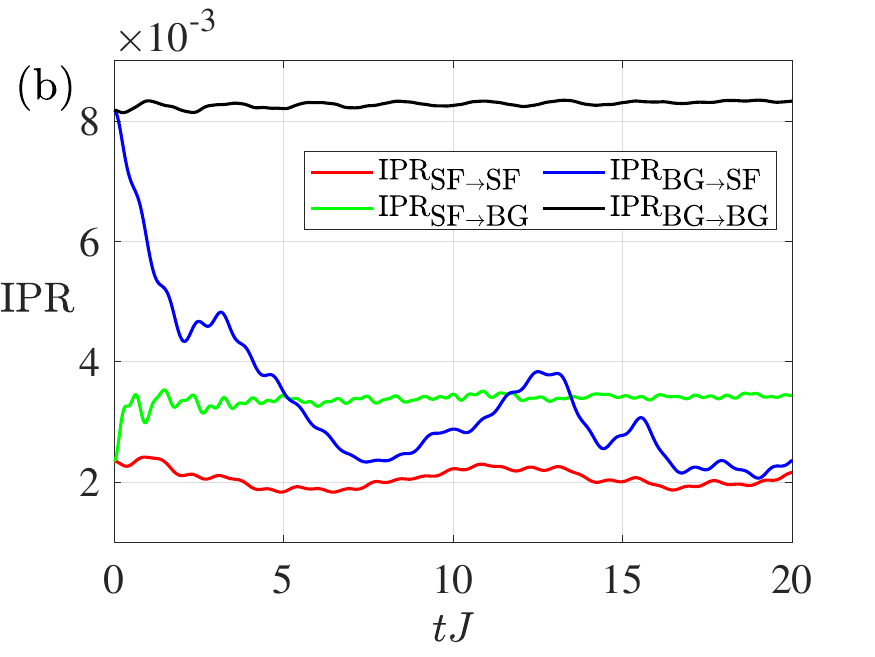}\includegraphics[totalheight=1.7in]{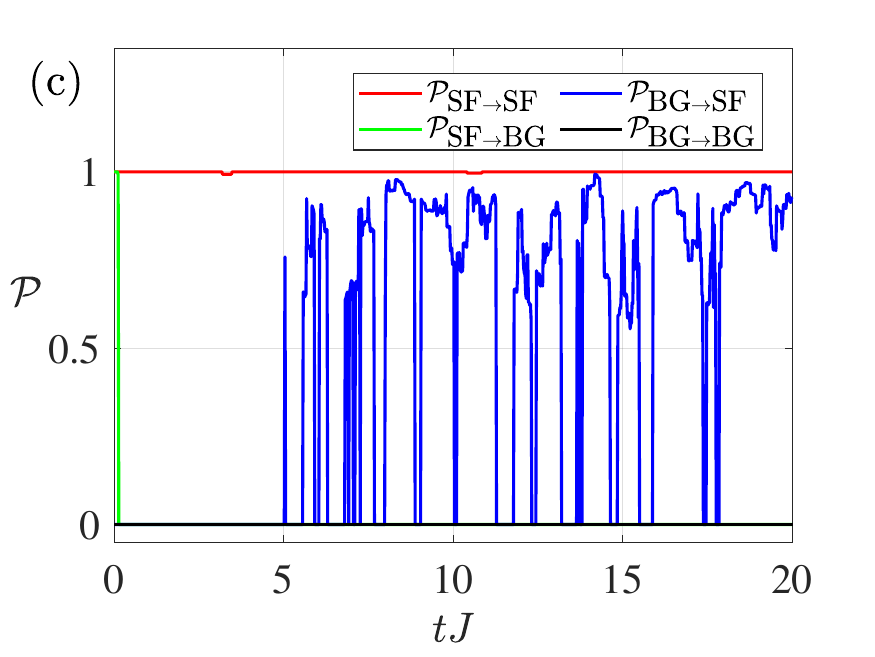}
\par\end{centering}
\begin{centering}
\includegraphics[scale=0.43]{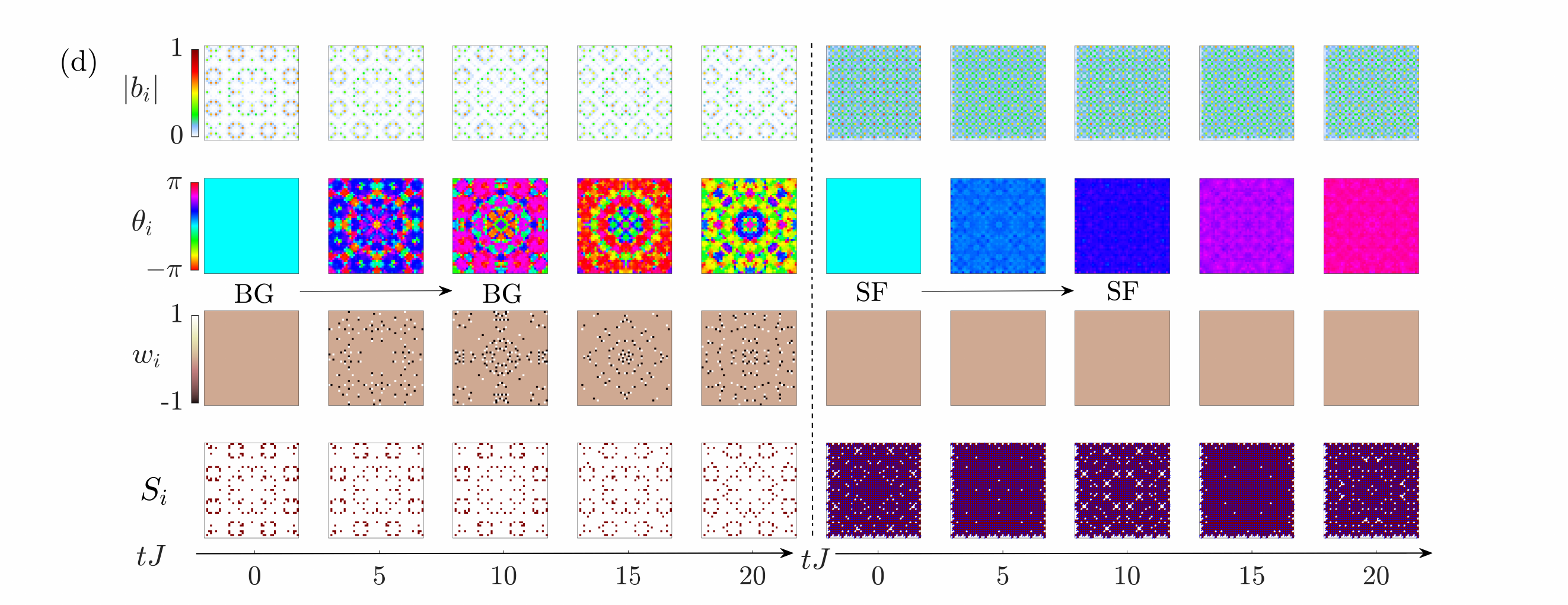}
\par\end{centering}
\begin{centering}
\includegraphics[scale=0.43]{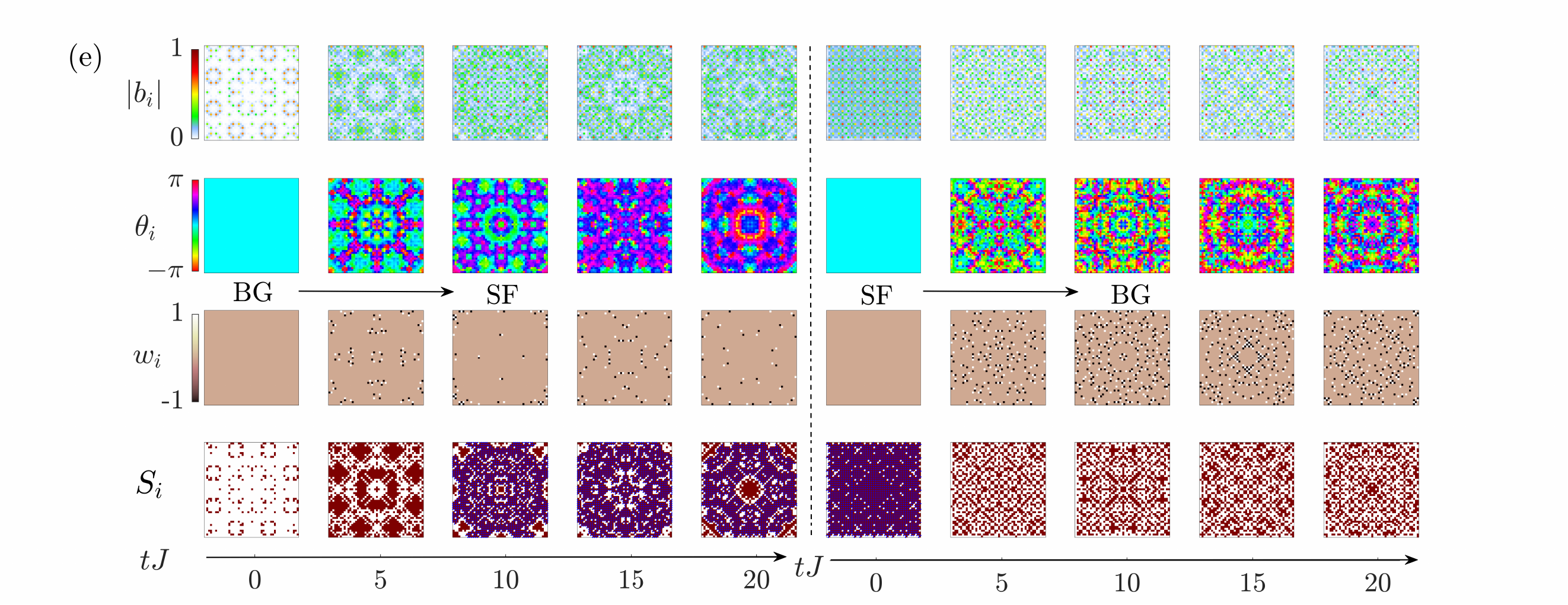}
\par\end{centering}
\caption{(a) Phase diagram of the system at $N=100$ with quench directions
marked with arrows. (b) Time dependence of the IPR of the corresponding
quench dynamics. (c) Time dependence of the percolation probability $\mathcal{P}$
of the corresponding quench dynamics. 
\textcolor{teal}{{} }Representative
real-space distributions of the amplitude of the SF order parameter
$\ensuremath{|\langle\hat{b}_{i}\rangle|}$, the phase $\ensuremath{\theta_{i}\equiv\arg\langle\hat{b}_{i}\rangle}$
of the SF order parameter, local winding number $w_{i}$ of the phase
field $\theta_{i}$, and the discrete field $S_{i}$ during (d) intraphase
and (e) interphase quench processes indicated by the arrows in
(a). See text for more details.\label{fig:quench_dynamics}}
\end{figure*}

\subsection{Quench dynamics}

Thus far, we have focused on the static properties of the system.
Now let us turn to the dynamical properties; in particular, here we
shall focus on the quench dynamics of the system. Specifically, via
the time-dependent Gutzwiller approach \citep{Schir0_2010_prl,Bloch_2008_rmp,Zakrzewski_2005_pra}
(see Appendix C for technique details of the approach), we investigate
quench dynamics of two different types, namely, intraphase quench
and interphase quench. 

Figure~\ref{fig:quench_dynamics}(d) shows the representative distributions
of the system during the intraphase quenches, where the system remains
within the same phase (e.g., from the BG or SF phase to the BG or SF phase).
Neither the time evolution of the IPR nor the percolation probability
$\mathcal{P}$ exhibits significant changes {[}see Figs.~\ref{fig:quench_dynamics}(b) and \ref{fig:quench_dynamics}(c){]}.
Likewise, the typical real-space distributions of the amplitude of
the SF order parameter and the discrete field at different time instants
show no significant variations {[}see Fig.~\ref{fig:quench_dynamics}(d){]}.
However, we do notice the distribution of the local winding number
$\ensuremath{w_{i}}$, defined for each plaquette of the lattice
as the winding of the phase field $\ensuremath{\theta_{i}\equiv\arg\langle\hat{b}_{i}\rangle}$
associated with the SF order parameter, i.e., $\ensuremath{w_{i}\equiv[(\theta_{i_{x}+1,i_{y}}-\theta_{i_{x},i_{y}})+(\theta_{i_{x}+1,i_{y}+1}-\theta_{i_{x}+1,i_{y}})+(\theta_{i_{x},i_{y}+1}-\theta_{i_{x}+1,i_{y}+1})+(\theta_{i_{x},i_{y}}-\theta_{i_{x},i_{y}+1})]/2\pi}$, undergoes
noticeable changes during the BG to BG quench. In this case, vortex
pairs with opposite charges, indicated by the nonzero local winding
number $\ensuremath{w_{i}}$, exhibit creation and annihilation dynamics.
This vortex behavior can be attributed to the reorganization of the
particle density distribution of the system during the quench dynamics,
in which the locations of sites with vanishing particle density---fertile
ground for the creation of vortex pairs---vary over time.

For interphase quenches, in which the system undergoes a transition
between different phases (e.g., from SF to BG or vice versa), we
observe more significant qualitative changes. Specifically, for these
quenches, the initial state corresponds to the ground-state configuration
at the two red dots in Fig.~\ref{fig:quench_dynamics}(a) with $(U/J=2,M_{r}/J=5)$
in the SF phase and $(U/J=2,M_{r}/J=10)$ in the BG phase. In the
quench from the SF phase to the BG phase, the percolation probability
$\mathcal{P}$ {[}illustrated by the green line in Fig.~\ref{fig:quench_dynamics}(c){]}
rapidly decreases from nearly 1 to 0, indicating the swift transition
of the system from SF to BG. This sharp drop in $\mathcal{P}$
originates from the fragmentation of the dominant percolated SF cluster
into disconnected SF islands, a process driven by particle hopping
in the vicinity of the percolated path; accordingly, the timescale
of this drop is determined by the hopping amplitude. Once the system
reaches the BG phase, $\mathcal{P}$ remains at zero, reflecting that
the system remains in the BG phase. This behavior arises because the
BG phase consists of isolated SF islands, which restrict particle
movement to these disconnected regions. As a result, the IPR remains
relatively constant with only minor fluctuations, corresponding to
slight variations in the spatial distribution of particles over time
{[}see the green line in Fig.~\ref{fig:quench_dynamics}(b){]}. It
is worth noting that the IPR does not fully rise to the value corresponding
to the final BG ground state. This difference originates from the
fact that the late-time state after the SF-to-BG quench is a nonequilibrium
state evolving from the SF ground state, whose conserved energy differs
from that of the BG ground state at the same parameters. The local-winding-number distribution $\ensuremath{w_{i}}$ also exhibits rapid
dynamics following the quench, characterized by an explosive generation
of vortex pairs {[}see the right panel in Fig.~\ref{fig:quench_dynamics}(e){]}.
This behavior signifies the breakdown of phase coherence on progressively
smaller length scales.\textcolor{red}{}

For the quench from the BG phase to the SF phase, the percolation
probability $\mathcal{P}$ fluctuates between zero and nonzero values
for times $tJ>5$, signaling the system's transition between the BG
and SF phases {[}see the blue line in Fig.~\ref{fig:quench_dynamics}(c){]}.
Unlike the quench from SF to BG, in which particles are confined to isolated
SF islands, in the BG-to-SF quench, particles can move freely between
SF clusters, allowing the formation of SF paths. However, at certain
moments, the system may lose its SF path and transition back to the
BG phase. This intermittent behavior gradually fades over time, and
as shown in Fig.~\ref{fig:quench_dynamics}, the periods during which
$\mathcal{P}=0$ become shorter, eventually leading the system to
settle into a steady SF phase. This trend is also reflected in the
time evolution of the IPR in Fig.~\ref{fig:quench_dynamics}(b).
Initially, in the BG phase, the particles are localized at specific
lattice sites, resulting in a relatively large IPR. As the system
evolves, particles diffuse through space, causing a gradual decrease
in the IPR. Although the IPR may occasionally increase, it generally
decreases over time, indicating that the system eventually transitions
to the SF phase. A similar pattern appears in the dynamics of the
local winding number distribution $\ensuremath{w_{i}}$ {[}see the
left panel in Fig.~\ref{fig:quench_dynamics}(e){]}. Over time, the
number of vortex pairs generally decreases, indicating the system's
progressive restoration of phase coherence across increasingly larger
length scales.

\section{CONCLUSIONS}

Our work demonstrated that ultracold bosonic systems in optical lattices
with an aperiodic external potential exhibit a rich interplay between
disorder and interaction effects. We showed that at low filling
factors, an increase in the on-site repulsive interaction triggers
a delocalization process that transforms a BG phase, characterized
by isolated and disconnected SF regions, into a robust SF phase with
a percolated network. This transition is clearly evidenced by the
evolution of the percolation probability and IPR. At higher filling
factors, we uncovered a striking reentrant behavior wherein the system
initially transitions from a BG to a SF phase as the interaction strength
increases, but further enhancement of the interaction causes the SF
network to fragment, causing the system to revert back to a BG phase. Moreover,
our analysis of quench dynamics revealed markedly different transient
behaviors for intraphase and interphase quenches, highlighting the
sensitive dependence of the system's dynamical evolution
on the initial state and interaction parameters. These findings not
only deepen our understanding of localization phenomena in interacting
Bose gases but also offer valuable insights for ongoing experiments
utilizing twisted bilayer and quasicrystalline optical lattice platforms,
where tuning disorder and interaction effects is pivotal for exploring
novel many-body phases.
\begin{acknowledgments}
This work is supported by the NKRDPC (Grant No.~2022YFA1405304),
the NSFC (Grant No.~12275089, No.~12574193, and No.~11904109), the Guangdong Basic
and Applied Basic Research Foundation (Grants No.~2024A1515010188,
No.~2023A1515012800, and No.~2021A1515010212), the Guangdong Provincial Key
Laboratory (Grant No.~2020B1212060066), and the Guangdong Provincial Quantum Science Strategic Initiative (Grant No.~GDZX2401002).
\end{acknowledgments}

\section{Data Availability}
The data that support the findings of this article are openly available \cite{data}.

\appendix

\section{TECHNICAL DETAILS FOR CALCULATING PERCOLATION PROBABILITY FROM A DISCRETE FIELD}

In this appendix, we provide technical details for calculating the
percolation probability $\mathcal{P}$ from the discrete field $S_{i}$.
The percolation probability $\mathcal{P}$ is defined in Ref. \citep{Johnstone_Duncan_2021_JPA} as 
\begin{equation}
\mathcal{P}=\frac{N_{\textrm{span}}}{N_{\varphi}},
\end{equation}
 where $N_{\textrm{span}}$ is the number of SF sites in a percolating
cluster and $N_{\varphi}$ is the total number of sites with finite
$\varphi$. In our numerical calculations, $N_{\textrm{span}}$ and
$N_{\varphi}$ are derived from the discrete field $S_{i}$ \citep{Niederle_2013_njp},
which can be constructed by the particle density distribution $\left\langle \hat{n}_{i}\right\rangle $
and SF order-parameter distribution $\langle\hat{b}_{i}\rangle$ described
in the main text. According to the definition of $S_{i}$ {[}see Eq.~(\ref{eq:Discrete_field_definition}){]},
$N_{\varphi}$ corresponds to the total number of sites where $S_{i}=1$,
i.e., $N_{\varphi}=\sum_{i}^{N_{\textrm{lat}}}S_{i}$. To calculate
$N_{\textrm{span}}$, it is necessary to identify the percolating
cluster. The criterion for percolation, as outlined in \citep{Barman_2013_epjb},
is the existence of at least one SF path that spans the entire system,
either from left to right or from top to bottom. Once such an SF path
is identified, the SF cluster that contains this SF path is considered
the percolating cluster. The total number of lattice sites in this
percolating cluster is then defined as $N_{\textrm{span}}$. Clearly,
as shown by the calculation of the percolation probability $\mathcal{P}$
described above, this quantity can be used to distinguish the SF phase
from the BG phase. In the BG phase, there are no percolating SF clusters,
which results in a percolation probability of zero. In contrast, in the
SF phase, the presence of percolating SF clusters leads to a nonzero
percolation probability.

\section{NUMERICAL RESULTS AT EVEN HIGHER FILLING FACTORS\label{appendix_high_filling}}

\begin{figure}[H]
\begin{centering}
\includegraphics[scale=0.5]{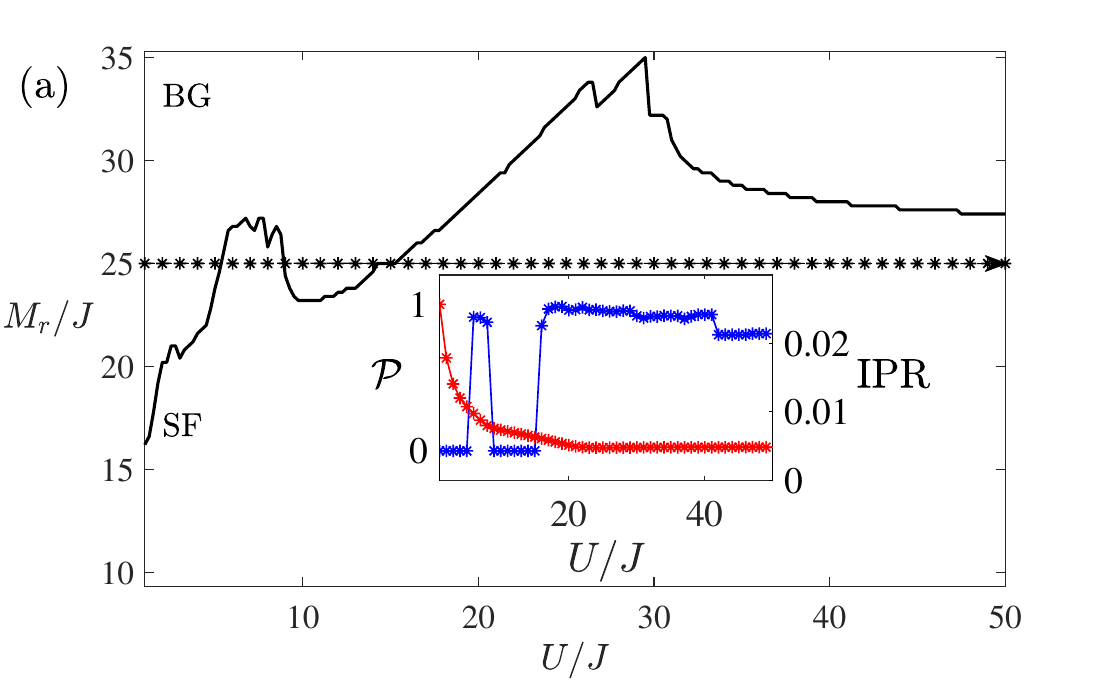}
\par\end{centering}
\begin{centering}
\includegraphics[scale=0.5]{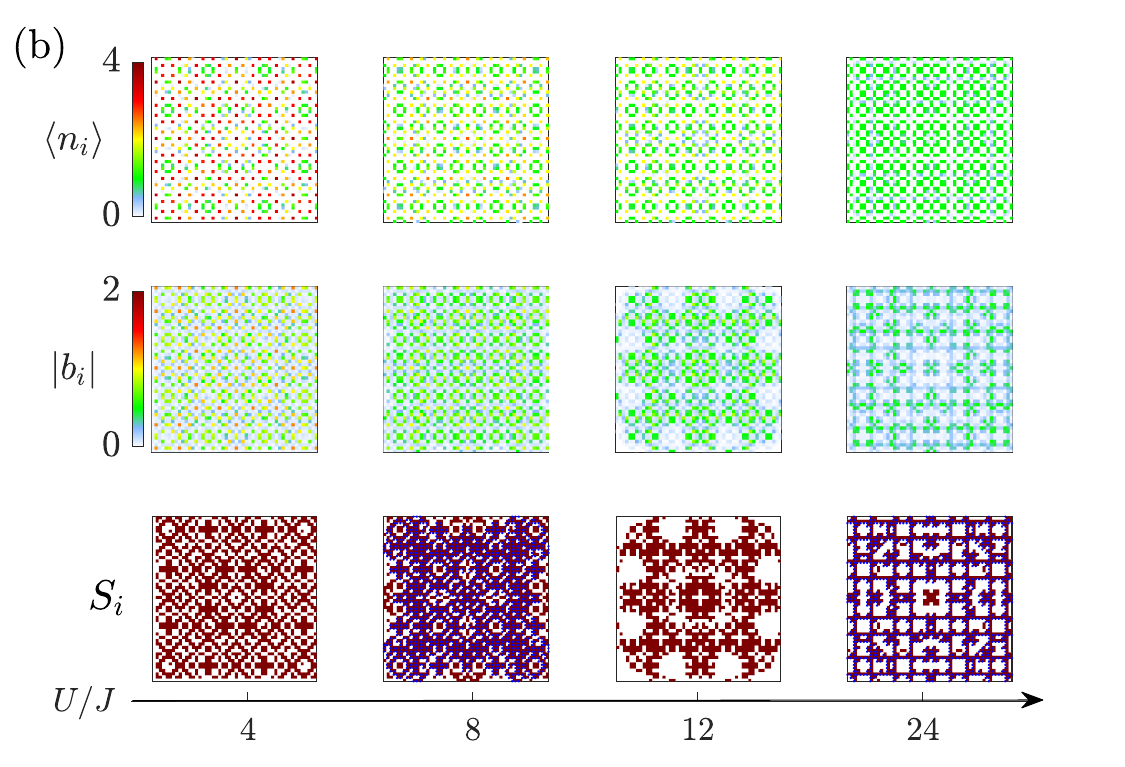}
\par\end{centering}
\caption{\label{fig:even higher filling}Phase diagram and representative real-space
distributions. (a) Phase diagram for a system of $1024$ particles
on a $51\times51$ square lattice. Inset: Dependence of the percolation
probability $\mathcal{P}$ (blue curve) and IPR (red curve) on $U/J$
for points along the dashed line indicated in the phase diagram. (b)
Real-space distributions of the particle density distribution $\ensuremath{\left\langle \hat{n}_{i}\right\rangle }$,
the amplitude of the SF order parameter $\ensuremath{|\langle\hat{b}_{i}\rangle|}$,
and the discrete field $S_{i}$ of four points along the dashed line
($M_{r}/J=25$). See text for more details.}
\end{figure}

As discussed in the main text, we hypothesize that multiple reentrance
phenomena could be observed in the system with an even higher particle
filling. Indeed, as anticipated, this behavior is demonstrated in
Fig.~\ref{fig:even higher filling}.

We plot the phase diagram for a particle filling of 1024, which
reveals two distinct peaks. Additionally, we includ a series of
real-space distribution diagrams showing the effect of varying the
interaction strength $U$ on the lattice, with the changes in $U$
indicated by the horizontal lines in the phase diagram. As shown in
the inset of the phase diagram, the percolation probability of the
ground state exhibits multiple transitions between zero and nonzero
values, corresponding to the system's behavior as it
transitions from the BG phase to the SF phase, then back to the BG
phase, and finally returns to the SF phase. The particle-density-distribution diagram Fig.~\ref{fig:even higher filling}(b) further
illustrates that, as the particle filling increases, the system undergoes
more significant changes due to the enhanced on-site interaction strength.
Moreover, these changes lead to the formation and annihilation of
SF path, creating a more complex phase transition process.

\begin{figure}[H]
\begin{centering}
\textcolor{red}{\includegraphics[scale=0.5]{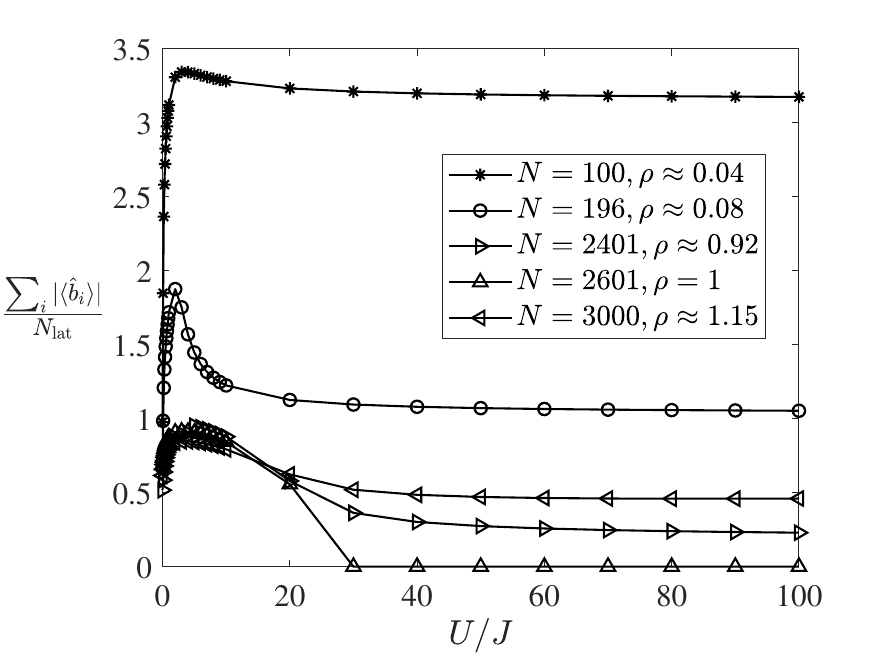}}
\par\end{centering}
\caption{\label{fig:Avg_SF_vs_U_at_diff_fillings}Dependence
of the spatially averaged SF order parameter $\sum_{i}|\langle\hat{b}_{i}\rangle|/N_{\mathrm{lat}}$
on the interaction strength at different filling factors. The aperiodic
potential strength is fixed at $M_{r}=7$, and the lattice size is
$L=51$ ($N_{\mathrm{lat}}=51\times51=2601$).}
\end{figure}

Moreover, at integer fillings, the standard localization
effect of strong interactions can also manifest in this system, driving
it into a Mott insulator phase. Figure~\ref{fig:Avg_SF_vs_U_at_diff_fillings}
shows the dependence of the spatially averaged superfluid (SF) order
parameter $\sum_{i}|\langle\hat{b}_{i}\rangle|/N_{\mathrm{lat}}$,
on the interaction strength $U$ for the different fillings. At integer
filling $\rho=1$, the SF order parameter vanishes at sufficiently
large $U/J$, clearly signaling the emergence of a Mott insulator
phase.

\section{Time dependent Gutzwiller approach}

In this appendix, we present the technical details of the time-dependent
Gutzwiller approach \citep{Schir0_2010_prl,Bloch_2008_rmp,Zakrzewski_2005_pra}.
This approach is built on top of the static Gutzwiller approach. The
static Gutzwiller method is employed to determine the ground state
by minimizing the total energy of the system, given by $E\left(\left\{ C_{i,n}\right\} \right)=\left\langle \textrm{GW}\right|\hat{H}\left|\textrm{GW}\right\rangle $,
while the dynamic Gutzwiller method minimizes the action to simulate
the system's evolution. Specifically, by applying the variational
principle, we can write the action of the system as $S=\int dt\left[-i\left\langle \textrm{GW}\left(t\right)\right|\partial_{t}\left|\textrm{GW}\left(t\right)\right\rangle +\left\langle \textrm{GW}\left(t\right)\right|\hat{H}\left|\textrm{GW}\left(t\right)\right\rangle \right]$.
Variation with respect to the coefficients $C_{i,n}\left(t\right)$
yields the equation of motion:
\begin{align}
	i\partial_{t}C_{i,n}\left(t\right) & =C_{i,n}\left(t\right)\left[\frac{U}{2}n\left(n-1\right)+M_{i}n\right]\nonumber \\
	& -J\left[C_{i,n+1}\sqrt{n+1}\left(\sum_{j}\Phi_{j}^{*}\right)\right.\\
	& \left.+C_{i,n-1}\sqrt{n}\left(\sum_{j}\Phi_{j}\right)\right]\nonumber 
\end{align}
where $\sum_{j}\Phi_{j}=\sum_{j}\sum_{n=0}^{n_{\textrm{max}}-1}C_{j,n+1}C_{j,n}^{*}\sqrt{n+1}$
and $\sum_{j}$ denotes the sum over neighboring lattice points. 

Time-dependent Gutzwiller method automatically conserves the expectation
value total particle number $\ensuremath{\left\langle \hat{N}\right\rangle }$
as long as $\ensuremath{[\hat{H},\hat{N}]=0}$. This can be illustrated
by considering a generic Gutzwiller state at time $\ensuremath{t}\ensuremath{\left|\textrm{GW}\left(t\right)\right\rangle }$.
After a time step $dt$, the state becomes:
\begin{equation}
\left|\textrm{GW}\left(t+dt\right)\right\rangle =\left|\textrm{GW}\left(t\right)\right\rangle -idt\hat{H}\left|\textrm{GW}\left(t\right)\right\rangle 
\end{equation}
The expectation value of the total particle number at the next time
step is
\begin{align}
\left\langle \hat{N}\left(t+dt\right)\right\rangle  & =\left\langle \textrm{GW}\left(t+dt\right)\right|\hat{N}\left|\textrm{GW}\left(t+dt\right)\right\rangle \nonumber \\
 & =\left\langle \hat{N}\left(t\right)\right\rangle +idt\left\langle \textrm{GW}\left(t\right)\right|\left[\hat{H},\hat{N}\right]\left|\textrm{GW}\left(t\right)\right\rangle \\
 & +\mathcal{O}\left(dt^{2}\right).\nonumber 
\end{align}
indicating $\ensuremath{\left\langle \hat{N}\right\rangle }$ is conserved
during evolution as long as $\ensuremath{[\hat{H},\hat{N}]=0}$.

\appendix


\begin{thebibliography}{82}%
	\makeatletter
	\providecommand \@ifxundefined [1]{%
		\@ifx{#1\undefined}
	}%
	\providecommand \@ifnum [1]{%
		\ifnum #1\expandafter \@firstoftwo
		\else \expandafter \@secondoftwo
		\fi
	}%
	\providecommand \@ifx [1]{%
		\ifx #1\expandafter \@firstoftwo
		\else \expandafter \@secondoftwo
		\fi
	}%
	\providecommand \natexlab [1]{#1}%
	\providecommand \enquote  [1]{``#1''}%
	\providecommand \bibnamefont  [1]{#1}%
	\providecommand \bibfnamefont [1]{#1}%
	\providecommand \citenamefont [1]{#1}%
	\providecommand \href@noop [0]{\@secondoftwo}%
	\providecommand \href [0]{\begingroup \@sanitize@url \@href}%
	\providecommand \@href[1]{\@@startlink{#1}\@@href}%
	\providecommand \@@href[1]{\endgroup#1\@@endlink}%
	\providecommand \@sanitize@url [0]{\catcode `\\12\catcode `\$12\catcode
		`\&12\catcode `\#12\catcode `\^12\catcode `\_12\catcode `\%12\relax}%
	\providecommand \@@startlink[1]{}%
	\providecommand \@@endlink[0]{}%
	\providecommand \url  [0]{\begingroup\@sanitize@url \@url }%
	\providecommand \@url [1]{\endgroup\@href {#1}{\urlprefix }}%
	\providecommand \urlprefix  [0]{URL }%
	\providecommand \Eprint [0]{\href }%
	\providecommand \doibase [0]{https://doi.org/}%
	\providecommand \selectlanguage [0]{\@gobble}%
	\providecommand \bibinfo  [0]{\@secondoftwo}%
	\providecommand \bibfield  [0]{\@secondoftwo}%
	\providecommand \translation [1]{[#1]}%
	\providecommand \BibitemOpen [0]{}%
	\providecommand \bibitemStop [0]{}%
	\providecommand \bibitemNoStop [0]{.\EOS\space}%
	\providecommand \EOS [0]{\spacefactor3000\relax}%
	\providecommand \BibitemShut  [1]{\csname bibitem#1\endcsname}%
	\let\auto@bib@innerbib\@empty
	%</preamble>
	\bibitem [{\citenamefont {Cao}\ \emph {et~al.}(2018{\natexlab{a}})\citenamefont
		{Cao}, \citenamefont {Fatemi}, \citenamefont {Demir}, \citenamefont {Fang},
		\citenamefont {Tomarken}, \citenamefont {Luo}, \citenamefont
		{{Sanchez-Yamagishi}}, \citenamefont {Watanabe}, \citenamefont {Taniguchi},
		\citenamefont {Kaxiras}, \citenamefont {Ashoori},\ and\ \citenamefont
		{{Jarillo-Herrero}}}]{cao_Nature_2018_Correlated_Insulator}%
	\BibitemOpen
	\bibfield  {author} {\bibinfo {author} {\bibfnamefont {Y.}~\bibnamefont
			{Cao}}, \bibinfo {author} {\bibfnamefont {V.}~\bibnamefont {Fatemi}},
		\bibinfo {author} {\bibfnamefont {A.}~\bibnamefont {Demir}}, \bibinfo
		{author} {\bibfnamefont {S.}~\bibnamefont {Fang}}, \bibinfo {author}
		{\bibfnamefont {S.~L.}\ \bibnamefont {Tomarken}}, \bibinfo {author}
		{\bibfnamefont {J.~Y.}\ \bibnamefont {Luo}}, \bibinfo {author} {\bibfnamefont
			{J.~D.}\ \bibnamefont {{Sanchez-Yamagishi}}}, \bibinfo {author}
		{\bibfnamefont {K.}~\bibnamefont {Watanabe}}, \bibinfo {author}
		{\bibfnamefont {T.}~\bibnamefont {Taniguchi}}, \bibinfo {author}
		{\bibfnamefont {E.}~\bibnamefont {Kaxiras}}, \bibinfo {author} {\bibfnamefont
			{R.~C.}\ \bibnamefont {Ashoori}},\ and\ \bibinfo {author} {\bibfnamefont
			{P.}~\bibnamefont {{Jarillo-Herrero}}},\ }\bibfield  {title} {\bibinfo
		{title} {Correlated insulator behaviour at half-filling in magic-angle
			graphene superlattices},\ }\href@noop {} {\bibfield  {journal} {\bibinfo
			{journal} {Nature}\ }\textbf {\bibinfo {volume} {556}},\ \bibinfo {pages}
		{80} (\bibinfo {year} {2018}{\natexlab{a}})}\BibitemShut {NoStop}%
	\bibitem [{\citenamefont {Li}\ \emph {et~al.}(2021{\natexlab{a}})\citenamefont
		{Li}, \citenamefont {Jiang}, \citenamefont {Li}, \citenamefont {Zhang},
		\citenamefont {Kang}, \citenamefont {Zhu}, \citenamefont {Watanabe},
		\citenamefont {Taniguchi}, \citenamefont {Chowdhury}, \citenamefont {Fu},
		\citenamefont {Shan},\ and\ \citenamefont
		{Mak}}]{li_Nature_2021_Continuous_Mott}%
	\BibitemOpen
	\bibfield  {author} {\bibinfo {author} {\bibfnamefont {T.}~\bibnamefont
			{Li}}, \bibinfo {author} {\bibfnamefont {S.}~\bibnamefont {Jiang}}, \bibinfo
		{author} {\bibfnamefont {L.}~\bibnamefont {Li}}, \bibinfo {author}
		{\bibfnamefont {Y.}~\bibnamefont {Zhang}}, \bibinfo {author} {\bibfnamefont
			{K.}~\bibnamefont {Kang}}, \bibinfo {author} {\bibfnamefont {J.}~\bibnamefont
			{Zhu}}, \bibinfo {author} {\bibfnamefont {K.}~\bibnamefont {Watanabe}},
		\bibinfo {author} {\bibfnamefont {T.}~\bibnamefont {Taniguchi}}, \bibinfo
		{author} {\bibfnamefont {D.}~\bibnamefont {Chowdhury}}, \bibinfo {author}
		{\bibfnamefont {L.}~\bibnamefont {Fu}}, \bibinfo {author} {\bibfnamefont
			{J.}~\bibnamefont {Shan}},\ and\ \bibinfo {author} {\bibfnamefont {K.~F.}\
			\bibnamefont {Mak}},\ }\bibfield  {title} {\bibinfo {title} {Continuous
			{Mott} transition in semiconductor moir\'e superlattices},\ }\href@noop {}
	{\bibfield  {journal} {\bibinfo  {journal} {Nature}\ }\textbf {\bibinfo
			{volume} {597}},\ \bibinfo {pages} {350} (\bibinfo {year}
		{2021}{\natexlab{a}})}\BibitemShut {NoStop}%
	\bibitem [{\citenamefont {Cao}\ \emph {et~al.}(2018{\natexlab{b}})\citenamefont
		{Cao}, \citenamefont {Fatemi}, \citenamefont {Fang}, \citenamefont
		{Watanabe}, \citenamefont {Taniguchi}, \citenamefont {Kaxiras},\ and\
		\citenamefont {{Jarillo-Herrero}}}]{cao_Nature_2018_Unconventional_SC}%
	\BibitemOpen
	\bibfield  {author} {\bibinfo {author} {\bibfnamefont {Y.}~\bibnamefont
			{Cao}}, \bibinfo {author} {\bibfnamefont {V.}~\bibnamefont {Fatemi}},
		\bibinfo {author} {\bibfnamefont {S.}~\bibnamefont {Fang}}, \bibinfo {author}
		{\bibfnamefont {K.}~\bibnamefont {Watanabe}}, \bibinfo {author}
		{\bibfnamefont {T.}~\bibnamefont {Taniguchi}}, \bibinfo {author}
		{\bibfnamefont {E.}~\bibnamefont {Kaxiras}},\ and\ \bibinfo {author}
		{\bibfnamefont {P.}~\bibnamefont {{Jarillo-Herrero}}},\ }\bibfield  {title}
	{\bibinfo {title} {Unconventional superconductivity in magic-angle graphene
			superlattices},\ }\href@noop {} {\bibfield  {journal} {\bibinfo  {journal}
			{Nature}\ }\textbf {\bibinfo {volume} {556}},\ \bibinfo {pages} {43}
		(\bibinfo {year} {2018}{\natexlab{b}})}\BibitemShut {NoStop}%
	\bibitem [{\citenamefont {Gong}\ \emph {et~al.}(2017)\citenamefont {Gong},
		\citenamefont {Li}, \citenamefont {Li}, \citenamefont {Ji}, \citenamefont
		{Stern}, \citenamefont {Xia}, \citenamefont {Cao}, \citenamefont {Bao},
		\citenamefont {Wang}, \citenamefont {Wang}, \citenamefont {Qiu},
		\citenamefont {Cava}, \citenamefont {Louie}, \citenamefont {Xia},\ and\
		\citenamefont {Zhang}}]{gong_Nature_2017}%
	\BibitemOpen
	\bibfield  {author} {\bibinfo {author} {\bibfnamefont {C.}~\bibnamefont
			{Gong}}, \bibinfo {author} {\bibfnamefont {L.}~\bibnamefont {Li}}, \bibinfo
		{author} {\bibfnamefont {Z.}~\bibnamefont {Li}}, \bibinfo {author}
		{\bibfnamefont {H.}~\bibnamefont {Ji}}, \bibinfo {author} {\bibfnamefont
			{A.}~\bibnamefont {Stern}}, \bibinfo {author} {\bibfnamefont
			{Y.}~\bibnamefont {Xia}}, \bibinfo {author} {\bibfnamefont {T.}~\bibnamefont
			{Cao}}, \bibinfo {author} {\bibfnamefont {W.}~\bibnamefont {Bao}}, \bibinfo
		{author} {\bibfnamefont {C.}~\bibnamefont {Wang}}, \bibinfo {author}
		{\bibfnamefont {Y.}~\bibnamefont {Wang}}, \bibinfo {author} {\bibfnamefont
			{Z.~Q.}\ \bibnamefont {Qiu}}, \bibinfo {author} {\bibfnamefont {R.~J.}\
			\bibnamefont {Cava}}, \bibinfo {author} {\bibfnamefont {S.~G.}\ \bibnamefont
			{Louie}}, \bibinfo {author} {\bibfnamefont {J.}~\bibnamefont {Xia}},\ and\
		\bibinfo {author} {\bibfnamefont {X.}~\bibnamefont {Zhang}},\ }\bibfield
	{title} {\bibinfo {title} {Discovery of intrinsic ferromagnetism in
			two-dimensional van der {{Waals}} crystals},\ }\href@noop {} {\bibfield
		{journal} {\bibinfo  {journal} {Nature}\ }\textbf {\bibinfo {volume} {546}},\
		\bibinfo {pages} {265} (\bibinfo {year} {2017})}\BibitemShut {NoStop}%
	\bibitem [{\citenamefont {Huang}\ \emph {et~al.}(2017)\citenamefont {Huang},
		\citenamefont {Clark}, \citenamefont {{Navarro-Moratalla}}, \citenamefont
		{Klein}, \citenamefont {Cheng}, \citenamefont {Seyler}, \citenamefont
		{Zhong}, \citenamefont {Schmidgall}, \citenamefont {McGuire}, \citenamefont
		{Cobden}, \citenamefont {Yao}, \citenamefont {Xiao}, \citenamefont
		{{Jarillo-Herrero}},\ and\ \citenamefont {Xu}}]{huang_Nature_2017}%
	\BibitemOpen
	\bibfield  {author} {\bibinfo {author} {\bibfnamefont {B.}~\bibnamefont
			{Huang}}, \bibinfo {author} {\bibfnamefont {G.}~\bibnamefont {Clark}},
		\bibinfo {author} {\bibfnamefont {E.}~\bibnamefont {{Navarro-Moratalla}}},
		\bibinfo {author} {\bibfnamefont {D.~R.}\ \bibnamefont {Klein}}, \bibinfo
		{author} {\bibfnamefont {R.}~\bibnamefont {Cheng}}, \bibinfo {author}
		{\bibfnamefont {K.~L.}\ \bibnamefont {Seyler}}, \bibinfo {author}
		{\bibfnamefont {D.}~\bibnamefont {Zhong}}, \bibinfo {author} {\bibfnamefont
			{E.}~\bibnamefont {Schmidgall}}, \bibinfo {author} {\bibfnamefont {M.~A.}\
			\bibnamefont {McGuire}}, \bibinfo {author} {\bibfnamefont {D.~H.}\
			\bibnamefont {Cobden}}, \bibinfo {author} {\bibfnamefont {W.}~\bibnamefont
			{Yao}}, \bibinfo {author} {\bibfnamefont {D.}~\bibnamefont {Xiao}}, \bibinfo
		{author} {\bibfnamefont {P.}~\bibnamefont {{Jarillo-Herrero}}},\ and\
		\bibinfo {author} {\bibfnamefont {X.}~\bibnamefont {Xu}},\ }\bibfield
	{title} {\bibinfo {title} {Layer-dependent ferromagnetism in a van der
			{{Waals}} crystal down to the monolayer limit},\ }\href@noop {} {\bibfield
		{journal} {\bibinfo  {journal} {Nature}\ }\textbf {\bibinfo {volume} {546}},\
		\bibinfo {pages} {270} (\bibinfo {year} {2017})}\BibitemShut {NoStop}%
	\bibitem [{\citenamefont {Chen}\ \emph {et~al.}(2020)\citenamefont {Chen},
		\citenamefont {Sharpe}, \citenamefont {Fox}, \citenamefont {Zhang},
		\citenamefont {Wang}, \citenamefont {Jiang}, \citenamefont {Lyu},
		\citenamefont {Li}, \citenamefont {Watanabe}, \citenamefont {Taniguchi},
		\citenamefont {Shi}, \citenamefont {Senthil}, \citenamefont
		{Goldhaber-Gordon}, \citenamefont {Zhang},\ and\ \citenamefont
		{Wang}}]{chen_Nature_2020}%
	\BibitemOpen
	\bibfield  {author} {\bibinfo {author} {\bibfnamefont {G.}~\bibnamefont
			{Chen}}, \bibinfo {author} {\bibfnamefont {A.~L.}\ \bibnamefont {Sharpe}},
		\bibinfo {author} {\bibfnamefont {E.~J.}\ \bibnamefont {Fox}}, \bibinfo
		{author} {\bibfnamefont {Y.-H.}\ \bibnamefont {Zhang}}, \bibinfo {author}
		{\bibfnamefont {S.}~\bibnamefont {Wang}}, \bibinfo {author} {\bibfnamefont
			{L.}~\bibnamefont {Jiang}}, \bibinfo {author} {\bibfnamefont
			{B.}~\bibnamefont {Lyu}}, \bibinfo {author} {\bibfnamefont {H.}~\bibnamefont
			{Li}}, \bibinfo {author} {\bibfnamefont {K.}~\bibnamefont {Watanabe}},
		\bibinfo {author} {\bibfnamefont {T.}~\bibnamefont {Taniguchi}}, \bibinfo
		{author} {\bibfnamefont {Z.}~\bibnamefont {Shi}}, \bibinfo {author}
		{\bibfnamefont {T.}~\bibnamefont {Senthil}}, \bibinfo {author} {\bibfnamefont
			{D.}~\bibnamefont {Goldhaber-Gordon}}, \bibinfo {author} {\bibfnamefont
			{Y.}~\bibnamefont {Zhang}},\ and\ \bibinfo {author} {\bibfnamefont
			{F.}~\bibnamefont {Wang}},\ }\bibfield  {title} {\bibinfo {title} {Tunable
			correlated {Chern} insulator and ferromagnetism in a moir\'e superlattice},\
	}\href@noop {} {\bibfield  {journal} {\bibinfo  {journal} {Nature}\ }\textbf
		{\bibinfo {volume} {579}},\ \bibinfo {pages} {56} (\bibinfo {year}
		{2020})}\BibitemShut {NoStop}%
	\bibitem [{\citenamefont {Li}\ \emph {et~al.}(2021{\natexlab{b}})\citenamefont
		{Li}, \citenamefont {Li}, \citenamefont {Regan}, \citenamefont {Wang},
		\citenamefont {Zhao}, \citenamefont {Kahn}, \citenamefont {Yumigeta},
		\citenamefont {Blei}, \citenamefont {Taniguchi}, \citenamefont {Watanabe},
		\citenamefont {Tongay}, \citenamefont {Zettl}, \citenamefont {Crommie},\ and\
		\citenamefont {Wang}}]{li_Hongyuan_Nature_2021}%
	\BibitemOpen
	\bibfield  {author} {\bibinfo {author} {\bibfnamefont {H.}~\bibnamefont
			{Li}}, \bibinfo {author} {\bibfnamefont {S.}~\bibnamefont {Li}}, \bibinfo
		{author} {\bibfnamefont {E.~C.}\ \bibnamefont {Regan}}, \bibinfo {author}
		{\bibfnamefont {D.}~\bibnamefont {Wang}}, \bibinfo {author} {\bibfnamefont
			{W.}~\bibnamefont {Zhao}}, \bibinfo {author} {\bibfnamefont {S.}~\bibnamefont
			{Kahn}}, \bibinfo {author} {\bibfnamefont {K.}~\bibnamefont {Yumigeta}},
		\bibinfo {author} {\bibfnamefont {M.}~\bibnamefont {Blei}}, \bibinfo {author}
		{\bibfnamefont {T.}~\bibnamefont {Taniguchi}}, \bibinfo {author}
		{\bibfnamefont {K.}~\bibnamefont {Watanabe}}, \bibinfo {author}
		{\bibfnamefont {S.}~\bibnamefont {Tongay}}, \bibinfo {author} {\bibfnamefont
			{A.}~\bibnamefont {Zettl}}, \bibinfo {author} {\bibfnamefont {M.~F.}\
			\bibnamefont {Crommie}},\ and\ \bibinfo {author} {\bibfnamefont
			{F.}~\bibnamefont {Wang}},\ }\bibfield  {title} {\bibinfo {title} {Imaging
			two-dimensional generalized {Wigner} crystals},\ }\href@noop {} {\bibfield
		{journal} {\bibinfo  {journal} {Nature}\ }\textbf {\bibinfo {volume} {597}},\
		\bibinfo {pages} {650} (\bibinfo {year} {2021}{\natexlab{b}})}\BibitemShut
	{NoStop}%
	\bibitem [{\citenamefont {Regan}\ \emph {et~al.}(2020)\citenamefont {Regan},
		\citenamefont {Wang}, \citenamefont {Jin}, \citenamefont {Bakti~Utama},
		\citenamefont {Gao}, \citenamefont {Wei}, \citenamefont {Zhao}, \citenamefont
		{Zhao}, \citenamefont {Zhang}, \citenamefont {Yumigeta}, \citenamefont
		{Blei}, \citenamefont {Carlstr{\"o}m}, \citenamefont {Watanabe},
		\citenamefont {Taniguchi}, \citenamefont {Tongay}, \citenamefont {Crommie},
		\citenamefont {Zettl},\ and\ \citenamefont {Wang}}]{regan_Nature_2020}%
	\BibitemOpen
	\bibfield  {author} {\bibinfo {author} {\bibfnamefont {E.~C.}\ \bibnamefont
			{Regan}}, \bibinfo {author} {\bibfnamefont {D.}~\bibnamefont {Wang}},
		\bibinfo {author} {\bibfnamefont {C.}~\bibnamefont {Jin}}, \bibinfo {author}
		{\bibfnamefont {M.~I.}\ \bibnamefont {Bakti~Utama}}, \bibinfo {author}
		{\bibfnamefont {B.}~\bibnamefont {Gao}}, \bibinfo {author} {\bibfnamefont
			{X.}~\bibnamefont {Wei}}, \bibinfo {author} {\bibfnamefont {S.}~\bibnamefont
			{Zhao}}, \bibinfo {author} {\bibfnamefont {W.}~\bibnamefont {Zhao}}, \bibinfo
		{author} {\bibfnamefont {Z.}~\bibnamefont {Zhang}}, \bibinfo {author}
		{\bibfnamefont {K.}~\bibnamefont {Yumigeta}}, \bibinfo {author}
		{\bibfnamefont {M.}~\bibnamefont {Blei}}, \bibinfo {author} {\bibfnamefont
			{J.~D.}\ \bibnamefont {Carlstr{\"o}m}}, \bibinfo {author} {\bibfnamefont
			{K.}~\bibnamefont {Watanabe}}, \bibinfo {author} {\bibfnamefont
			{T.}~\bibnamefont {Taniguchi}}, \bibinfo {author} {\bibfnamefont
			{S.}~\bibnamefont {Tongay}}, \bibinfo {author} {\bibfnamefont
			{M.}~\bibnamefont {Crommie}}, \bibinfo {author} {\bibfnamefont
			{A.}~\bibnamefont {Zettl}},\ and\ \bibinfo {author} {\bibfnamefont
			{F.}~\bibnamefont {Wang}},\ }\bibfield  {title} {\bibinfo {title} {Mott and
			generalized {{Wigner}} crystal states in {{WSe$_2$}}/{{WS$_2$}} moir\'e
			superlattices},\ }\href@noop {} {\bibfield  {journal} {\bibinfo  {journal}
			{Nature}\ }\textbf {\bibinfo {volume} {579}},\ \bibinfo {pages} {359}
		(\bibinfo {year} {2020})}\BibitemShut {NoStop}%
	\bibitem [{\citenamefont {Serlin}\ \emph {et~al.}(2020)\citenamefont {Serlin},
		\citenamefont {Tschirhart}, \citenamefont {Polshyn}, \citenamefont {Zhang},
		\citenamefont {Zhu}, \citenamefont {Watanabe}, \citenamefont {Taniguchi},
		\citenamefont {Balents},\ and\ \citenamefont {Young}}]{serlin_Science_2020}%
	\BibitemOpen
	\bibfield  {author} {\bibinfo {author} {\bibfnamefont {M.}~\bibnamefont
			{Serlin}}, \bibinfo {author} {\bibfnamefont {C.~L.}\ \bibnamefont
			{Tschirhart}}, \bibinfo {author} {\bibfnamefont {H.}~\bibnamefont {Polshyn}},
		\bibinfo {author} {\bibfnamefont {Y.}~\bibnamefont {Zhang}}, \bibinfo
		{author} {\bibfnamefont {J.}~\bibnamefont {Zhu}}, \bibinfo {author}
		{\bibfnamefont {K.}~\bibnamefont {Watanabe}}, \bibinfo {author}
		{\bibfnamefont {T.}~\bibnamefont {Taniguchi}}, \bibinfo {author}
		{\bibfnamefont {L.}~\bibnamefont {Balents}},\ and\ \bibinfo {author}
		{\bibfnamefont {A.~F.}\ \bibnamefont {Young}},\ }\bibfield  {title} {\bibinfo
		{title} {Intrinsic quantized anomalous {Hall} effect in a moir\'e
			heterostructure},\ }\href@noop {} {\bibfield  {journal} {\bibinfo  {journal}
			{Science}\ }\textbf {\bibinfo {volume} {367}},\ \bibinfo {pages} {900}
		(\bibinfo {year} {2020})}\BibitemShut {NoStop}%
	\bibitem [{\citenamefont {Li}\ \emph {et~al.}(2021{\natexlab{c}})\citenamefont
		{Li}, \citenamefont {Jiang}, \citenamefont {Shen}, \citenamefont {Zhang},
		\citenamefont {Li}, \citenamefont {Tao}, \citenamefont {Devakul},
		\citenamefont {Watanabe}, \citenamefont {Taniguchi}, \citenamefont {Fu},
		\citenamefont {Shan},\ and\ \citenamefont {Mak}}]{li_Nature_2021_QAH}%
	\BibitemOpen
	\bibfield  {author} {\bibinfo {author} {\bibfnamefont {T.}~\bibnamefont
			{Li}}, \bibinfo {author} {\bibfnamefont {S.}~\bibnamefont {Jiang}}, \bibinfo
		{author} {\bibfnamefont {B.}~\bibnamefont {Shen}}, \bibinfo {author}
		{\bibfnamefont {Y.}~\bibnamefont {Zhang}}, \bibinfo {author} {\bibfnamefont
			{L.}~\bibnamefont {Li}}, \bibinfo {author} {\bibfnamefont {Z.}~\bibnamefont
			{Tao}}, \bibinfo {author} {\bibfnamefont {T.}~\bibnamefont {Devakul}},
		\bibinfo {author} {\bibfnamefont {K.}~\bibnamefont {Watanabe}}, \bibinfo
		{author} {\bibfnamefont {T.}~\bibnamefont {Taniguchi}}, \bibinfo {author}
		{\bibfnamefont {L.}~\bibnamefont {Fu}}, \bibinfo {author} {\bibfnamefont
			{J.}~\bibnamefont {Shan}},\ and\ \bibinfo {author} {\bibfnamefont {K.~F.}\
			\bibnamefont {Mak}},\ }\bibfield  {title} {\bibinfo {title} {Quantum
			anomalous {Hall} effect from intertwined moir\'e bands},\ }\href@noop {}
	{\bibfield  {journal} {\bibinfo  {journal} {Nature}\ }\textbf {\bibinfo
			{volume} {600}},\ \bibinfo {pages} {641} (\bibinfo {year}
		{2021}{\natexlab{c}})}\BibitemShut {NoStop}%
	\bibitem [{\citenamefont {Zhou}\ \emph {et~al.}(2021)\citenamefont {Zhou},
		\citenamefont {Sung}, \citenamefont {Brutschea}, \citenamefont {Esterlis},
		\citenamefont {Wang}, \citenamefont {Scuri}, \citenamefont {Gelly},
		\citenamefont {Heo}, \citenamefont {Taniguchi}, \citenamefont {Watanabe},
		\citenamefont {Zar{\'A}nd}, \citenamefont {Lukin}, \citenamefont {Kim},
		\citenamefont {Demler},\ and\ \citenamefont {Park}}]{zhou_Nature_2021a}%
	\BibitemOpen
	\bibfield  {author} {\bibinfo {author} {\bibfnamefont {Y.}~\bibnamefont
			{Zhou}}, \bibinfo {author} {\bibfnamefont {J.}~\bibnamefont {Sung}}, \bibinfo
		{author} {\bibfnamefont {E.}~\bibnamefont {Brutschea}}, \bibinfo {author}
		{\bibfnamefont {I.}~\bibnamefont {Esterlis}}, \bibinfo {author}
		{\bibfnamefont {Y.}~\bibnamefont {Wang}}, \bibinfo {author} {\bibfnamefont
			{G.}~\bibnamefont {Scuri}}, \bibinfo {author} {\bibfnamefont {R.~J.}\
			\bibnamefont {Gelly}}, \bibinfo {author} {\bibfnamefont {H.}~\bibnamefont
			{Heo}}, \bibinfo {author} {\bibfnamefont {T.}~\bibnamefont {Taniguchi}},
		\bibinfo {author} {\bibfnamefont {K.}~\bibnamefont {Watanabe}}, \bibinfo
		{author} {\bibfnamefont {G.}~\bibnamefont {Zar{\'A}nd}}, \bibinfo {author}
		{\bibfnamefont {M.~D.}\ \bibnamefont {Lukin}}, \bibinfo {author}
		{\bibfnamefont {P.}~\bibnamefont {Kim}}, \bibinfo {author} {\bibfnamefont
			{E.}~\bibnamefont {Demler}},\ and\ \bibinfo {author} {\bibfnamefont
			{H.}~\bibnamefont {Park}},\ }\bibfield  {title} {\bibinfo {title} {Bilayer
			{Wigner} crystals in a transition metal dichalcogenide heterostructure},\
	}\href@noop {} {\bibfield  {journal} {\bibinfo  {journal} {Nature}\ }\textbf
		{\bibinfo {volume} {595}},\ \bibinfo {pages} {48} (\bibinfo {year}
		{2021})}\BibitemShut {NoStop}%
	\bibitem [{\citenamefont {Cai}\ \emph {et~al.}(2023)\citenamefont {Cai},
		\citenamefont {Anderson}, \citenamefont {Wang}, \citenamefont {Zhang},
		\citenamefont {Liu}, \citenamefont {Holtzmann}, \citenamefont {Zhang},
		\citenamefont {Fan}, \citenamefont {Taniguchi}, \citenamefont {Watanabe},
		\citenamefont {Ran}, \citenamefont {Cao}, \citenamefont {Fu}, \citenamefont
		{Xiao}, \citenamefont {Yao},\ and\ \citenamefont
		{Xu}}]{Cai_Nature_2023_add_1}%
	\BibitemOpen
	\bibfield  {author} {\bibinfo {author} {\bibfnamefont {J.}~\bibnamefont
			{Cai}}, \bibinfo {author} {\bibfnamefont {E.}~\bibnamefont {Anderson}},
		\bibinfo {author} {\bibfnamefont {C.}~\bibnamefont {Wang}}, \bibinfo {author}
		{\bibfnamefont {X.}~\bibnamefont {Zhang}}, \bibinfo {author} {\bibfnamefont
			{X.}~\bibnamefont {Liu}}, \bibinfo {author} {\bibfnamefont {W.}~\bibnamefont
			{Holtzmann}}, \bibinfo {author} {\bibfnamefont {Y.}~\bibnamefont {Zhang}},
		\bibinfo {author} {\bibfnamefont {F.}~\bibnamefont {Fan}}, \bibinfo {author}
		{\bibfnamefont {T.}~\bibnamefont {Taniguchi}}, \bibinfo {author}
		{\bibfnamefont {K.}~\bibnamefont {Watanabe}}, \bibinfo {author}
		{\bibfnamefont {Y.}~\bibnamefont {Ran}}, \bibinfo {author} {\bibfnamefont
			{T.}~\bibnamefont {Cao}}, \bibinfo {author} {\bibfnamefont {L.}~\bibnamefont
			{Fu}}, \bibinfo {author} {\bibfnamefont {D.}~\bibnamefont {Xiao}}, \bibinfo
		{author} {\bibfnamefont {W.}~\bibnamefont {Yao}},\ and\ \bibinfo {author}
		{\bibfnamefont {X.}~\bibnamefont {Xu}},\ }\bibfield  {title} {\bibinfo
		{title} {Signatures of fractional quantum anomalous {Hall} states in twisted
			{MoTe$_2$}},\ }\href@noop {} {\bibfield  {journal} {\bibinfo  {journal}
			{Nature}\ }\textbf {\bibinfo {volume} {622}},\ \bibinfo {pages} {63}
		(\bibinfo {year} {2023})}\BibitemShut {NoStop}%
	\bibitem [{\citenamefont {Park}\ \emph {et~al.}(2023)\citenamefont {Park},
		\citenamefont {Cai}, \citenamefont {Anderson}, \citenamefont {Zhang},
		\citenamefont {Zhu}, \citenamefont {Liu}, \citenamefont {Wang}, \citenamefont
		{Holtzmann}, \citenamefont {Hu}, \citenamefont {Liu}, \citenamefont
		{Taniguchi}, \citenamefont {Watanabe}, \citenamefont {Chu}, \citenamefont
		{Cao}, \citenamefont {Fu}, \citenamefont {Yao}, \citenamefont {Chang},
		\citenamefont {Cobden}, \citenamefont {Xiao},\ and\ \citenamefont
		{Xu}}]{Park_Nature_2023_add_2}%
	\BibitemOpen
	\bibfield  {author} {\bibinfo {author} {\bibfnamefont {H.}~\bibnamefont
			{Park}}, \bibinfo {author} {\bibfnamefont {J.}~\bibnamefont {Cai}}, \bibinfo
		{author} {\bibfnamefont {E.}~\bibnamefont {Anderson}}, \bibinfo {author}
		{\bibfnamefont {Y.}~\bibnamefont {Zhang}}, \bibinfo {author} {\bibfnamefont
			{J.}~\bibnamefont {Zhu}}, \bibinfo {author} {\bibfnamefont {X.}~\bibnamefont
			{Liu}}, \bibinfo {author} {\bibfnamefont {C.}~\bibnamefont {Wang}}, \bibinfo
		{author} {\bibfnamefont {W.}~\bibnamefont {Holtzmann}}, \bibinfo {author}
		{\bibfnamefont {C.}~\bibnamefont {Hu}}, \bibinfo {author} {\bibfnamefont
			{Z.}~\bibnamefont {Liu}}, \bibinfo {author} {\bibfnamefont {T.}~\bibnamefont
			{Taniguchi}}, \bibinfo {author} {\bibfnamefont {K.}~\bibnamefont {Watanabe}},
		\bibinfo {author} {\bibfnamefont {J.-H.}\ \bibnamefont {Chu}}, \bibinfo
		{author} {\bibfnamefont {T.}~\bibnamefont {Cao}}, \bibinfo {author}
		{\bibfnamefont {L.}~\bibnamefont {Fu}}, \bibinfo {author} {\bibfnamefont
			{W.}~\bibnamefont {Yao}}, \bibinfo {author} {\bibfnamefont {C.-Z.}\
			\bibnamefont {Chang}}, \bibinfo {author} {\bibfnamefont {D.}~\bibnamefont
			{Cobden}}, \bibinfo {author} {\bibfnamefont {D.}~\bibnamefont {Xiao}},\ and\
		\bibinfo {author} {\bibfnamefont {X.}~\bibnamefont {Xu}},\ }\bibfield
	{title} {\bibinfo {title} {Observation of fractionally quantized anomalous
			{Hall} effect},\ }\href@noop {} {\bibfield  {journal} {\bibinfo  {journal}
			{Nature}\ }\textbf {\bibinfo {volume} {622}},\ \bibinfo {pages} {74}
		(\bibinfo {year} {2023})}\BibitemShut {NoStop}%
	\bibitem [{\citenamefont {Zeng}\ \emph {et~al.}(2023)\citenamefont {Zeng},
		\citenamefont {Xia}, \citenamefont {Kang}, \citenamefont {Zhu}, \citenamefont
		{Kn{\"u}ppel}, \citenamefont {Vaswani}, \citenamefont {Watanabe},
		\citenamefont {Taniguchi}, \citenamefont {Mak},\ and\ \citenamefont
		{Shan}}]{Zeng_2023_add_3}%
	\BibitemOpen
	\bibfield  {author} {\bibinfo {author} {\bibfnamefont {Y.}~\bibnamefont
			{Zeng}}, \bibinfo {author} {\bibfnamefont {Z.}~\bibnamefont {Xia}}, \bibinfo
		{author} {\bibfnamefont {K.}~\bibnamefont {Kang}}, \bibinfo {author}
		{\bibfnamefont {J.}~\bibnamefont {Zhu}}, \bibinfo {author} {\bibfnamefont
			{P.}~\bibnamefont {Kn{\"u}ppel}}, \bibinfo {author} {\bibfnamefont
			{C.}~\bibnamefont {Vaswani}}, \bibinfo {author} {\bibfnamefont
			{K.}~\bibnamefont {Watanabe}}, \bibinfo {author} {\bibfnamefont
			{T.}~\bibnamefont {Taniguchi}}, \bibinfo {author} {\bibfnamefont {K.~F.}\
			\bibnamefont {Mak}},\ and\ \bibinfo {author} {\bibfnamefont {J.}~\bibnamefont
			{Shan}},\ }\bibfield  {title} {\bibinfo {title} {Thermodynamic evidence of
			fractional {Chern} insulator in moir{\'e} {MoTe$_2$}},\ }\href@noop {}
	{\bibfield  {journal} {\bibinfo  {journal} {Nature}\ }\textbf {\bibinfo
			{volume} {622}},\ \bibinfo {pages} {69} (\bibinfo {year} {2023})}\BibitemShut
	{NoStop}%
	\bibitem [{\citenamefont {Xu}\ \emph {et~al.}(2023)\citenamefont {Xu},
		\citenamefont {Sun}, \citenamefont {Jia}, \citenamefont {Liu}, \citenamefont
		{Xu}, \citenamefont {Li}, \citenamefont {Gu}, \citenamefont {Watanabe},
		\citenamefont {Taniguchi}, \citenamefont {Tong}, \citenamefont {Jia},
		\citenamefont {Shi}, \citenamefont {Jiang}, \citenamefont {Zhang},
		\citenamefont {Liu},\ and\ \citenamefont {Li}}]{Xu_2023_PRX_add_4}%
	\BibitemOpen
	\bibfield  {author} {\bibinfo {author} {\bibfnamefont {F.}~\bibnamefont
			{Xu}}, \bibinfo {author} {\bibfnamefont {Z.}~\bibnamefont {Sun}}, \bibinfo
		{author} {\bibfnamefont {T.}~\bibnamefont {Jia}}, \bibinfo {author}
		{\bibfnamefont {C.}~\bibnamefont {Liu}}, \bibinfo {author} {\bibfnamefont
			{C.}~\bibnamefont {Xu}}, \bibinfo {author} {\bibfnamefont {C.}~\bibnamefont
			{Li}}, \bibinfo {author} {\bibfnamefont {Y.}~\bibnamefont {Gu}}, \bibinfo
		{author} {\bibfnamefont {K.}~\bibnamefont {Watanabe}}, \bibinfo {author}
		{\bibfnamefont {T.}~\bibnamefont {Taniguchi}}, \bibinfo {author}
		{\bibfnamefont {B.}~\bibnamefont {Tong}}, \bibinfo {author} {\bibfnamefont
			{J.}~\bibnamefont {Jia}}, \bibinfo {author} {\bibfnamefont {Z.}~\bibnamefont
			{Shi}}, \bibinfo {author} {\bibfnamefont {S.}~\bibnamefont {Jiang}}, \bibinfo
		{author} {\bibfnamefont {Y.}~\bibnamefont {Zhang}}, \bibinfo {author}
		{\bibfnamefont {X.}~\bibnamefont {Liu}},\ and\ \bibinfo {author}
		{\bibfnamefont {T.}~\bibnamefont {Li}},\ }\bibfield  {title} {\bibinfo
		{title} {Observation of integer and fractional quantum anomalous {{Hall}}
			effects in twisted bilayer {{MoTe$_2$}}},\ }\href@noop {} {\bibfield
		{journal} {\bibinfo  {journal} {Phys. Rev. X}\ }\textbf {\bibinfo {volume}
			{13}},\ \bibinfo {pages} {031037} (\bibinfo {year} {2023})}\BibitemShut
	{NoStop}%
	\bibitem [{\citenamefont {Lu}\ \emph {et~al.}(2024)\citenamefont {Lu},
		\citenamefont {Han}, \citenamefont {Yao}, \citenamefont {Reddy},
		\citenamefont {Yang}, \citenamefont {Seo}, \citenamefont {Watanabe},
		\citenamefont {Taniguchi}, \citenamefont {Fu},\ and\ \citenamefont
		{Ju}}]{Lu_2024_add_5}%
	\BibitemOpen
	\bibfield  {author} {\bibinfo {author} {\bibfnamefont {Z.}~\bibnamefont
			{Lu}}, \bibinfo {author} {\bibfnamefont {T.}~\bibnamefont {Han}}, \bibinfo
		{author} {\bibfnamefont {Y.}~\bibnamefont {Yao}}, \bibinfo {author}
		{\bibfnamefont {A.~P.}\ \bibnamefont {Reddy}}, \bibinfo {author}
		{\bibfnamefont {J.}~\bibnamefont {Yang}}, \bibinfo {author} {\bibfnamefont
			{J.}~\bibnamefont {Seo}}, \bibinfo {author} {\bibfnamefont {K.}~\bibnamefont
			{Watanabe}}, \bibinfo {author} {\bibfnamefont {T.}~\bibnamefont {Taniguchi}},
		\bibinfo {author} {\bibfnamefont {L.}~\bibnamefont {Fu}},\ and\ \bibinfo
		{author} {\bibfnamefont {L.}~\bibnamefont {Ju}},\ }\bibfield  {title}
	{\bibinfo {title} {Fractional quantum anomalous {{Hall}} effect in multilayer
			graphene},\ }\href@noop {} {\bibfield  {journal} {\bibinfo  {journal}
			{Nature}\ }\textbf {\bibinfo {volume} {626}},\ \bibinfo {pages} {759}
		(\bibinfo {year} {2024})}\BibitemShut {NoStop}%
	\bibitem [{\citenamefont {Shechtman}\ \emph {et~al.}(1984)\citenamefont
		{Shechtman}, \citenamefont {Blech}, \citenamefont {Gratias},\ and\
		\citenamefont {Cahn}}]{Shechtman_Cahn_1984_PRL}%
	\BibitemOpen
	\bibfield  {author} {\bibinfo {author} {\bibfnamefont {D.}~\bibnamefont
			{Shechtman}}, \bibinfo {author} {\bibfnamefont {I.}~\bibnamefont {Blech}},
		\bibinfo {author} {\bibfnamefont {D.}~\bibnamefont {Gratias}},\ and\ \bibinfo
		{author} {\bibfnamefont {J.~W.}\ \bibnamefont {Cahn}},\ }\bibfield  {title}
	{\bibinfo {title} {Metallic phase with long-range orientational order and no
			translational symmetry},\ }\href@noop {} {\bibfield  {journal} {\bibinfo
			{journal} {Phys. Rev. Lett.}\ }\textbf {\bibinfo {volume} {53}},\ \bibinfo
		{pages} {1951} (\bibinfo {year} {1984})}\BibitemShut {NoStop}%
	\bibitem [{\citenamefont {Uri}\ \emph {et~al.}(2023)\citenamefont {Uri},
		\citenamefont {de~la Barrera}, \citenamefont {Randeria}, \citenamefont
		{Rodan-Legrain}, \citenamefont {Devakul}, \citenamefont {Crowley},
		\citenamefont {Paul}, \citenamefont {Watanabe}, \citenamefont {Taniguchi},
		\citenamefont {Lifshitz}, \citenamefont {Fu}, \citenamefont {Ashoori},\ and\
		\citenamefont {Jarillo-Herrero}}]{Uri_Jarillo-Herrero_2023_Nature}%
	\BibitemOpen
	\bibfield  {author} {\bibinfo {author} {\bibfnamefont {A.}~\bibnamefont
			{Uri}}, \bibinfo {author} {\bibfnamefont {S.~C.}\ \bibnamefont {de~la
				Barrera}}, \bibinfo {author} {\bibfnamefont {M.~T.}\ \bibnamefont
			{Randeria}}, \bibinfo {author} {\bibfnamefont {D.}~\bibnamefont
			{Rodan-Legrain}}, \bibinfo {author} {\bibfnamefont {T.}~\bibnamefont
			{Devakul}}, \bibinfo {author} {\bibfnamefont {P.~J.~D.}\ \bibnamefont
			{Crowley}}, \bibinfo {author} {\bibfnamefont {N.}~\bibnamefont {Paul}},
		\bibinfo {author} {\bibfnamefont {K.}~\bibnamefont {Watanabe}}, \bibinfo
		{author} {\bibfnamefont {T.}~\bibnamefont {Taniguchi}}, \bibinfo {author}
		{\bibfnamefont {R.}~\bibnamefont {Lifshitz}}, \bibinfo {author}
		{\bibfnamefont {L.}~\bibnamefont {Fu}}, \bibinfo {author} {\bibfnamefont
			{R.~C.}\ \bibnamefont {Ashoori}},\ and\ \bibinfo {author} {\bibfnamefont
			{P.}~\bibnamefont {Jarillo-Herrero}},\ }\bibfield  {title} {\bibinfo {title}
		{Superconductivity and strong interactions in a tunable
			moir{\'e}quasicrystal},\ }\href@noop {} {\bibfield  {journal} {\bibinfo
			{journal} {Nature (London)}\ }\textbf {\bibinfo {volume} {620}},\ \bibinfo
		{pages} {762} (\bibinfo {year} {2023})}\BibitemShut {NoStop}%
	\bibitem [{\citenamefont {Li}\ \emph {et~al.}(2024)\citenamefont {Li},
		\citenamefont {Zhang}, \citenamefont {Ha}, \citenamefont {Lin}, \citenamefont
		{Dong}, \citenamefont {Gao}, \citenamefont {Liu}, \citenamefont {Liu},
		\citenamefont {Ryu}, \citenamefont {Kim}, \citenamefont {Jozwiak},
		\citenamefont {Bostwick}, \citenamefont {Watanabe}, \citenamefont
		{Taniguchi}, \citenamefont {Kousa}, \citenamefont {Li}, \citenamefont
		{Rotenberg}, \citenamefont {Khalaf}, \citenamefont {Robinson}, \citenamefont
		{Giustino},\ and\ \citenamefont {Shih}}]{Li_Shih_2024_Nature}%
	\BibitemOpen
	\bibfield  {author} {\bibinfo {author} {\bibfnamefont {Y.}~\bibnamefont
			{Li}}, \bibinfo {author} {\bibfnamefont {F.}~\bibnamefont {Zhang}}, \bibinfo
		{author} {\bibfnamefont {V.-A.}\ \bibnamefont {Ha}}, \bibinfo {author}
		{\bibfnamefont {Y.-C.}\ \bibnamefont {Lin}}, \bibinfo {author} {\bibfnamefont
			{C.}~\bibnamefont {Dong}}, \bibinfo {author} {\bibfnamefont {Q.}~\bibnamefont
			{Gao}}, \bibinfo {author} {\bibfnamefont {Z.}~\bibnamefont {Liu}}, \bibinfo
		{author} {\bibfnamefont {X.}~\bibnamefont {Liu}}, \bibinfo {author}
		{\bibfnamefont {S.~H.}\ \bibnamefont {Ryu}}, \bibinfo {author} {\bibfnamefont
			{H.}~\bibnamefont {Kim}}, \bibinfo {author} {\bibfnamefont {C.}~\bibnamefont
			{Jozwiak}}, \bibinfo {author} {\bibfnamefont {A.}~\bibnamefont {Bostwick}},
		\bibinfo {author} {\bibfnamefont {K.}~\bibnamefont {Watanabe}}, \bibinfo
		{author} {\bibfnamefont {T.}~\bibnamefont {Taniguchi}}, \bibinfo {author}
		{\bibfnamefont {B.}~\bibnamefont {Kousa}}, \bibinfo {author} {\bibfnamefont
			{X.}~\bibnamefont {Li}}, \bibinfo {author} {\bibfnamefont {E.}~\bibnamefont
			{Rotenberg}}, \bibinfo {author} {\bibfnamefont {E.}~\bibnamefont {Khalaf}},
		\bibinfo {author} {\bibfnamefont {J.~A.}\ \bibnamefont {Robinson}}, \bibinfo
		{author} {\bibfnamefont {F.}~\bibnamefont {Giustino}},\ and\ \bibinfo
		{author} {\bibfnamefont {C.-K.}\ \bibnamefont {Shih}},\ }\bibfield  {title}
	{\bibinfo {title} {Tuning commensurability in twisted van der {Waals}
			bilayers},\ }\href@noop {} {\bibfield  {journal} {\bibinfo  {journal} {Nature
				(London)}\ }\textbf {\bibinfo {volume} {625}},\ \bibinfo {pages} {494}
		(\bibinfo {year} {2024})}\BibitemShut {NoStop}%
	\bibitem [{\citenamefont {Park}\ \emph {et~al.}(2019)\citenamefont {Park},
		\citenamefont {Kim},\ and\ \citenamefont {Lee}}]{Park_Lee_2019_PRB}%
	\BibitemOpen
	\bibfield  {author} {\bibinfo {author} {\bibfnamefont {M.~J.}\ \bibnamefont
			{Park}}, \bibinfo {author} {\bibfnamefont {H.~S.}\ \bibnamefont {Kim}},\ and\
		\bibinfo {author} {\bibfnamefont {S.}~\bibnamefont {Lee}},\ }\bibfield
	{title} {\bibinfo {title} {Emergent localization in dodecagonal bilayer
			quasicrystals},\ }\href@noop {} {\bibfield  {journal} {\bibinfo  {journal}
			{Phys. Rev. B}\ }\textbf {\bibinfo {volume} {99}},\ \bibinfo {pages} {245401}
		(\bibinfo {year} {2019})}\BibitemShut {NoStop}%
	\bibitem [{\citenamefont {Huang}\ and\ \citenamefont
		{Liu}(2019)}]{Huang_Liu_2019_PRB}%
	\BibitemOpen
	\bibfield  {author} {\bibinfo {author} {\bibfnamefont {B.}~\bibnamefont
			{Huang}}\ and\ \bibinfo {author} {\bibfnamefont {W.~V.}\ \bibnamefont
			{Liu}},\ }\bibfield  {title} {\bibinfo {title} {Moir\'e localization in
			two-dimensional quasiperiodic systems},\ }\href@noop {} {\bibfield  {journal}
		{\bibinfo  {journal} {Phys. Rev. B}\ }\textbf {\bibinfo {volume} {100}},\
		\bibinfo {pages} {144202} (\bibinfo {year} {2019})}\BibitemShut {NoStop}%
	\bibitem [{\citenamefont {Anderson}(1958)}]{Anderson_1958_PR}%
	\BibitemOpen
	\bibfield  {author} {\bibinfo {author} {\bibfnamefont {P.~W.}\ \bibnamefont
			{Anderson}},\ }\bibfield  {title} {\bibinfo {title} {Absence of diffusion in
			certain random lattices},\ }\href@noop {} {\bibfield  {journal} {\bibinfo
			{journal} {Phys. Rev.}\ }\textbf {\bibinfo {volume} {109}},\ \bibinfo {pages}
		{1492} (\bibinfo {year} {1958})}\BibitemShut {NoStop}%
	\bibitem [{\citenamefont {Aubry}\ and\ \citenamefont
		{Andr\'{e}}(1980)}]{Aubry_Andre_1980_AIPS}%
	\BibitemOpen
	\bibfield  {author} {\bibinfo {author} {\bibfnamefont {S.}~\bibnamefont
			{Aubry}}\ and\ \bibinfo {author} {\bibfnamefont {G.}~\bibnamefont
			{Andr\'{e}}},\ }\bibfield  {title} {\bibinfo {title} {Analyticity breaking
			and {Anderson} localization in incommensurate lattices},\ }\href@noop {}
	{\bibfield  {journal} {\bibinfo  {journal} {Ann. Israel Phys. Soc.}\ }\textbf
		{\bibinfo {volume} {3}},\ \bibinfo {pages} {18} (\bibinfo {year}
		{1980})}\BibitemShut {NoStop}%
	\bibitem [{\citenamefont {Roati}\ \emph {et~al.}(2008)\citenamefont {Roati},
		\citenamefont {D'Errico}, \citenamefont {Fallani}, \citenamefont {Fattori},
		\citenamefont {Fort}, \citenamefont {Zaccanti}, \citenamefont {Modugno},
		\citenamefont {Modugno},\ and\ \citenamefont
		{Inguscio}}]{Roati_Inguscio_2008_Nature}%
	\BibitemOpen
	\bibfield  {author} {\bibinfo {author} {\bibfnamefont {G.}~\bibnamefont
			{Roati}}, \bibinfo {author} {\bibfnamefont {C.}~\bibnamefont {D'Errico}},
		\bibinfo {author} {\bibfnamefont {L.}~\bibnamefont {Fallani}}, \bibinfo
		{author} {\bibfnamefont {M.}~\bibnamefont {Fattori}}, \bibinfo {author}
		{\bibfnamefont {C.}~\bibnamefont {Fort}}, \bibinfo {author} {\bibfnamefont
			{M.}~\bibnamefont {Zaccanti}}, \bibinfo {author} {\bibfnamefont
			{G.}~\bibnamefont {Modugno}}, \bibinfo {author} {\bibfnamefont
			{M.}~\bibnamefont {Modugno}},\ and\ \bibinfo {author} {\bibfnamefont
			{M.}~\bibnamefont {Inguscio}},\ }\bibfield  {title} {\bibinfo {title}
		{Anderson localization of a non-interacting {Bose}--{Einstein} condensate},\
	}\href@noop {} {\bibfield  {journal} {\bibinfo  {journal} {Nature}\ }\textbf
		{\bibinfo {volume} {453}},\ \bibinfo {pages} {895} (\bibinfo {year}
		{2008})}\BibitemShut {NoStop}%
	\bibitem [{\citenamefont {Lahini}\ \emph {et~al.}(2009)\citenamefont {Lahini},
		\citenamefont {Pugatch}, \citenamefont {Pozzi}, \citenamefont {Sorel},
		\citenamefont {Morandotti}, \citenamefont {Davidson},\ and\ \citenamefont
		{Silberberg}}]{Lahini_Silberberg_2009_PRL}%
	\BibitemOpen
	\bibfield  {author} {\bibinfo {author} {\bibfnamefont {Y.}~\bibnamefont
			{Lahini}}, \bibinfo {author} {\bibfnamefont {R.}~\bibnamefont {Pugatch}},
		\bibinfo {author} {\bibfnamefont {F.}~\bibnamefont {Pozzi}}, \bibinfo
		{author} {\bibfnamefont {M.}~\bibnamefont {Sorel}}, \bibinfo {author}
		{\bibfnamefont {R.}~\bibnamefont {Morandotti}}, \bibinfo {author}
		{\bibfnamefont {N.}~\bibnamefont {Davidson}},\ and\ \bibinfo {author}
		{\bibfnamefont {Y.}~\bibnamefont {Silberberg}},\ }\bibfield  {title}
	{\bibinfo {title} {Observation of a localization transition in quasiperiodic
			photonic lattices},\ }\href@noop {} {\bibfield  {journal} {\bibinfo
			{journal} {Phys. Rev. Lett.}\ }\textbf {\bibinfo {volume} {103}},\ \bibinfo
		{pages} {013901} (\bibinfo {year} {2009})}\BibitemShut {NoStop}%
	\bibitem [{\citenamefont {Abrahams}\ \emph {et~al.}(1979)\citenamefont
		{Abrahams}, \citenamefont {Anderson}, \citenamefont {Licciardello},\ and\
		\citenamefont {Ramakrishnan}}]{Abrahams_Ramakrishnan_1979_PRL}%
	\BibitemOpen
	\bibfield  {author} {\bibinfo {author} {\bibfnamefont {E.}~\bibnamefont
			{Abrahams}}, \bibinfo {author} {\bibfnamefont {P.~W.}\ \bibnamefont
			{Anderson}}, \bibinfo {author} {\bibfnamefont {D.~C.}\ \bibnamefont
			{Licciardello}},\ and\ \bibinfo {author} {\bibfnamefont {T.~V.}\ \bibnamefont
			{Ramakrishnan}},\ }\bibfield  {title} {\bibinfo {title} {Scaling theory of
			localization: Absence of quantum diffusion in two dimensions},\ }\href@noop
	{} {\bibfield  {journal} {\bibinfo  {journal} {Phys. Rev. Lett.}\ }\textbf
		{\bibinfo {volume} {42}},\ \bibinfo {pages} {673} (\bibinfo {year}
		{1979})}\BibitemShut {NoStop}%
	\bibitem [{\citenamefont {Iyer}\ \emph {et~al.}(2013)\citenamefont {Iyer},
		\citenamefont {Oganesyan}, \citenamefont {Refael},\ and\ \citenamefont
		{Huse}}]{Iyer_Huse_2013_PRB}%
	\BibitemOpen
	\bibfield  {author} {\bibinfo {author} {\bibfnamefont {S.}~\bibnamefont
			{Iyer}}, \bibinfo {author} {\bibfnamefont {V.}~\bibnamefont {Oganesyan}},
		\bibinfo {author} {\bibfnamefont {G.}~\bibnamefont {Refael}},\ and\ \bibinfo
		{author} {\bibfnamefont {D.~A.}\ \bibnamefont {Huse}},\ }\bibfield  {title}
	{\bibinfo {title} {Many-body localization in a quasiperiodic system},\
	}\href@noop {} {\bibfield  {journal} {\bibinfo  {journal} {Phys. Rev. B}\
		}\textbf {\bibinfo {volume} {87}},\ \bibinfo {pages} {134202} (\bibinfo
		{year} {2013})}\BibitemShut {NoStop}%
	\bibitem [{\citenamefont {Mondaini}\ and\ \citenamefont
		{Rigol}(2015)}]{Mondaini_Rigol_2015_PRA}%
	\BibitemOpen
	\bibfield  {author} {\bibinfo {author} {\bibfnamefont {R.}~\bibnamefont
			{Mondaini}}\ and\ \bibinfo {author} {\bibfnamefont {M.}~\bibnamefont
			{Rigol}},\ }\bibfield  {title} {\bibinfo {title} {Many-body localization and
			thermalization in disordered {Hubbard} chains},\ }\href@noop {} {\bibfield
		{journal} {\bibinfo  {journal} {Phys. Rev. A}\ }\textbf {\bibinfo {volume}
			{92}},\ \bibinfo {pages} {041601} (\bibinfo {year} {2015})}\BibitemShut
	{NoStop}%
	\bibitem [{\citenamefont {Xu}\ \emph {et~al.}(2019)\citenamefont {Xu},
		\citenamefont {Li}, \citenamefont {Hsu}, \citenamefont {Swingle},\ and\
		\citenamefont {Das~Sarma}}]{Xu_DasSarma_2019_PRR}%
	\BibitemOpen
	\bibfield  {author} {\bibinfo {author} {\bibfnamefont {S.}~\bibnamefont
			{Xu}}, \bibinfo {author} {\bibfnamefont {X.}~\bibnamefont {Li}}, \bibinfo
		{author} {\bibfnamefont {Y.-T.}\ \bibnamefont {Hsu}}, \bibinfo {author}
		{\bibfnamefont {B.}~\bibnamefont {Swingle}},\ and\ \bibinfo {author}
		{\bibfnamefont {S.}~\bibnamefont {Das~Sarma}},\ }\bibfield  {title} {\bibinfo
		{title} {Butterfly effect in interacting {Aubry-Andr\'e} model:
			Thermalization, slow scrambling, and many-body localization},\ }\href@noop {}
	{\bibfield  {journal} {\bibinfo  {journal} {Phys. Rev. Res.}\ }\textbf
		{\bibinfo {volume} {1}},\ \bibinfo {pages} {032039(R)} (\bibinfo {year}
		{2019})}\BibitemShut {NoStop}%
	\bibitem [{\citenamefont {Vu}\ \emph {et~al.}(2022)\citenamefont {Vu},
		\citenamefont {Huang}, \citenamefont {Li},\ and\ \citenamefont
		{Das~Sarma}}]{Vu_DasSarma_2022_PRL}%
	\BibitemOpen
	\bibfield  {author} {\bibinfo {author} {\bibfnamefont {D.~D.}\ \bibnamefont
			{Vu}}, \bibinfo {author} {\bibfnamefont {K.}~\bibnamefont {Huang}}, \bibinfo
		{author} {\bibfnamefont {X.}~\bibnamefont {Li}},\ and\ \bibinfo {author}
		{\bibfnamefont {S.}~\bibnamefont {Das~Sarma}},\ }\bibfield  {title} {\bibinfo
		{title} {Fermionic many-body localization for random and quasiperiodic
			systems in the presence of short- and long-range interactions},\ }\href@noop
	{} {\bibfield  {journal} {\bibinfo  {journal} {Phys. Rev. Lett.}\ }\textbf
		{\bibinfo {volume} {128}},\ \bibinfo {pages} {146601} (\bibinfo {year}
		{2022})}\BibitemShut {NoStop}%
	\bibitem [{\citenamefont {Wang}\ \emph {et~al.}(2021)\citenamefont {Wang},
		\citenamefont {Cheng}, \citenamefont {Liu},\ and\ \citenamefont
		{Yu}}]{Wang_Yu_2021_PRL}%
	\BibitemOpen
	\bibfield  {author} {\bibinfo {author} {\bibfnamefont {Y.}~\bibnamefont
			{Wang}}, \bibinfo {author} {\bibfnamefont {C.}~\bibnamefont {Cheng}},
		\bibinfo {author} {\bibfnamefont {X.-J.}\ \bibnamefont {Liu}},\ and\ \bibinfo
		{author} {\bibfnamefont {D.}~\bibnamefont {Yu}},\ }\bibfield  {title}
	{\bibinfo {title} {Many-body critical phase: Extended and nonthermal},\
	}\href@noop {} {\bibfield  {journal} {\bibinfo  {journal} {Phys. Rev. Lett.}\
		}\textbf {\bibinfo {volume} {126}},\ \bibinfo {pages} {080602} (\bibinfo
		{year} {2021})}\BibitemShut {NoStop}%
	\bibitem [{\citenamefont {Vu}\ and\ \citenamefont
		{Das~Sarma}(2021)}]{Vu_DasSarma_2021_PRL}%
	\BibitemOpen
	\bibfield  {author} {\bibinfo {author} {\bibfnamefont {D.~D.}\ \bibnamefont
			{Vu}}\ and\ \bibinfo {author} {\bibfnamefont {S.}~\bibnamefont {Das~Sarma}},\
	}\bibfield  {title} {\bibinfo {title} {Moir\'e versus {Mott}:
			Incommensuration and interaction in one-dimensional bichromatic lattices},\
	}\href@noop {} {\bibfield  {journal} {\bibinfo  {journal} {Phys. Rev. Lett.}\
		}\textbf {\bibinfo {volume} {126}},\ \bibinfo {pages} {036803} (\bibinfo
		{year} {2021})}\BibitemShut {NoStop}%
	\bibitem [{\citenamefont {Gon{\c c}alves}\ \emph
		{et~al.}(2024{\natexlab{a}})\citenamefont {Gon{\c c}alves}, \citenamefont
		{Amorim}, \citenamefont {Riche}, \citenamefont {Castro},\ and\ \citenamefont
		{Ribeiro}}]{Goncalves_Ribeiro_2024_NatPhys}%
	\BibitemOpen
	\bibfield  {author} {\bibinfo {author} {\bibfnamefont {M.}~\bibnamefont
			{Gon{\c c}alves}}, \bibinfo {author} {\bibfnamefont {B.}~\bibnamefont
			{Amorim}}, \bibinfo {author} {\bibfnamefont {F.}~\bibnamefont {Riche}},
		\bibinfo {author} {\bibfnamefont {E.~V.}\ \bibnamefont {Castro}},\ and\
		\bibinfo {author} {\bibfnamefont {P.}~\bibnamefont {Ribeiro}},\ }\bibfield
	{title} {\bibinfo {title} {Incommensurability enabled quasi-fractal order in
			{1D} narrow-band moir{\'e}systems},\ }\href@noop {} {\bibfield  {journal}
		{\bibinfo  {journal} {Nat. Phys.}\ }\textbf {\bibinfo {volume} {20}},\
		\bibinfo {pages} {1933} (\bibinfo {year} {2024}{\natexlab{a}})}\BibitemShut
	{NoStop}%
	\bibitem [{\citenamefont {Kraus}\ \emph {et~al.}(2014)\citenamefont {Kraus},
		\citenamefont {Zilberberg},\ and\ \citenamefont
		{Berkovits}}]{Kraus_Berkovits_2014_PRB}%
	\BibitemOpen
	\bibfield  {author} {\bibinfo {author} {\bibfnamefont {Y.~E.}\ \bibnamefont
			{Kraus}}, \bibinfo {author} {\bibfnamefont {O.}~\bibnamefont {Zilberberg}},\
		and\ \bibinfo {author} {\bibfnamefont {R.}~\bibnamefont {Berkovits}},\
	}\bibfield  {title} {\bibinfo {title} {Enhanced compressibility due to
			repulsive interaction in the {Harper} model},\ }\href@noop {} {\bibfield
		{journal} {\bibinfo  {journal} {Phys. Rev. B}\ }\textbf {\bibinfo {volume}
			{89}},\ \bibinfo {pages} {161106(R)} (\bibinfo {year} {2014})}\BibitemShut
	{NoStop}%
	\bibitem [{\citenamefont {Cookmeyer}\ \emph {et~al.}(2020)\citenamefont
		{Cookmeyer}, \citenamefont {Motruk},\ and\ \citenamefont
		{Moore}}]{Cookmeyer_Moore_2020_PRB}%
	\BibitemOpen
	\bibfield  {author} {\bibinfo {author} {\bibfnamefont {T.}~\bibnamefont
			{Cookmeyer}}, \bibinfo {author} {\bibfnamefont {J.}~\bibnamefont {Motruk}},\
		and\ \bibinfo {author} {\bibfnamefont {J.~E.}\ \bibnamefont {Moore}},\
	}\bibfield  {title} {\bibinfo {title} {Critical properties of the
			ground-state localization-delocalization transition in the many-particle
			{Aubry-Andr\'e} model},\ }\href@noop {} {\bibfield  {journal} {\bibinfo
			{journal} {Phys. Rev. B}\ }\textbf {\bibinfo {volume} {101}},\ \bibinfo
		{pages} {174203} (\bibinfo {year} {2020})}\BibitemShut {NoStop}%
	\bibitem [{\citenamefont {Oliveira}\ \emph {et~al.}()\citenamefont {Oliveira},
		\citenamefont {Gon{\c c}alves}, \citenamefont {beiro}, \citenamefont
		{Castro},\ and\ \citenamefont {Amorim}}]{Oliveira_Amorim_2024_arXiv}%
	\BibitemOpen
	\bibfield  {author} {\bibinfo {author} {\bibfnamefont {R.}~\bibnamefont
			{Oliveira}}, \bibinfo {author} {\bibfnamefont {M.}~\bibnamefont {Gon{\c
					c}alves}}, \bibinfo {author} {\bibfnamefont {P.}~\bibnamefont {beiro}},
		\bibinfo {author} {\bibfnamefont {E.~V.}\ \bibnamefont {Castro}},\ and\
		\bibinfo {author} {\bibfnamefont {B.}~\bibnamefont {Amorim}},\ }\bibfield
	{title} {\bibinfo {title} {Incommensurability-induced enhancement of
			superconductivity in one dimensional critical systems},\ }\href@noop {}
	{\bibinfo  {journal} {arXiv:2303.17656}\ }\BibitemShut {NoStop}%
	\bibitem [{\citenamefont {Gon{\c c}alves}\ \emph
		{et~al.}(2024{\natexlab{b}})\citenamefont {Gon{\c c}alves}, \citenamefont
		{Pixley}, \citenamefont {Amorim}, \citenamefont {Castro},\ and\ \citenamefont
		{Ribeiro}}]{Gonccalves_Ribeiro_2024_PRB}%
	\BibitemOpen
	\bibfield  {journal} {  }\bibfield  {author} {\bibinfo {author} {\bibfnamefont
			{M.}~\bibnamefont {Gon{\c c}alves}}, \bibinfo {author} {\bibfnamefont
			{J.~H.}\ \bibnamefont {Pixley}}, \bibinfo {author} {\bibfnamefont
			{B.}~\bibnamefont {Amorim}}, \bibinfo {author} {\bibfnamefont {E.~V.}\
			\bibnamefont {Castro}},\ and\ \bibinfo {author} {\bibfnamefont
			{P.}~\bibnamefont {Ribeiro}},\ }\bibfield  {title} {\bibinfo {title}
		{Short-range interactions are irrelevant at the quasiperiodicity-driven
			{Luttinger} liquid to {Anderson} glass transition},\ }\href@noop {}
	{\bibfield  {journal} {\bibinfo  {journal} {Phys. Rev. B}\ }\textbf {\bibinfo
			{volume} {109}},\ \bibinfo {pages} {014211} (\bibinfo {year}
		{2024}{\natexlab{b}})}\BibitemShut {NoStop}%
	\bibitem [{\citenamefont {Giamarchi}\ and\ \citenamefont
		{Schulz}(1988)}]{Giamarchi_Schulz_1988_PRB}%
	\BibitemOpen
	\bibfield  {author} {\bibinfo {author} {\bibfnamefont {T.}~\bibnamefont
			{Giamarchi}}\ and\ \bibinfo {author} {\bibfnamefont {H.~J.}\ \bibnamefont
			{Schulz}},\ }\bibfield  {title} {\bibinfo {title} {Anderson localization and
			interactions in one-dimensional metals},\ }\href@noop {} {\bibfield
		{journal} {\bibinfo  {journal} {Phys. Rev. B}\ }\textbf {\bibinfo {volume}
			{37}},\ \bibinfo {pages} {325} (\bibinfo {year} {1988})}\BibitemShut
	{NoStop}%
	\bibitem [{\citenamefont {Fisher}\ \emph {et~al.}(1989)\citenamefont {Fisher},
		\citenamefont {Weichman}, \citenamefont {Grinstein},\ and\ \citenamefont
		{Fisher}}]{Fisher_Fisher_1989_PRB}%
	\BibitemOpen
	\bibfield  {author} {\bibinfo {author} {\bibfnamefont {M.~P.~A.}\
			\bibnamefont {Fisher}}, \bibinfo {author} {\bibfnamefont {P.~B.}\
			\bibnamefont {Weichman}}, \bibinfo {author} {\bibfnamefont {G.}~\bibnamefont
			{Grinstein}},\ and\ \bibinfo {author} {\bibfnamefont {D.~S.}\ \bibnamefont
			{Fisher}},\ }\bibfield  {title} {\bibinfo {title} {Boson localization and the
			superfluid-insulator transition},\ }\href@noop {} {\bibfield  {journal}
		{\bibinfo  {journal} {Phys. Rev. B}\ }\textbf {\bibinfo {volume} {40}},\
		\bibinfo {pages} {546} (\bibinfo {year} {1989})}\BibitemShut {NoStop}%
	\bibitem [{\citenamefont {Crowell}\ \emph {et~al.}(1997)\citenamefont
		{Crowell}, \citenamefont {Van~Keuls},\ and\ \citenamefont
		{Reppy}}]{Crowell_Reppy_1997_PRB}%
	\BibitemOpen
	\bibfield  {author} {\bibinfo {author} {\bibfnamefont {P.~A.}\ \bibnamefont
			{Crowell}}, \bibinfo {author} {\bibfnamefont {F.~W.}\ \bibnamefont
			{Van~Keuls}},\ and\ \bibinfo {author} {\bibfnamefont {J.~D.}\ \bibnamefont
			{Reppy}},\ }\bibfield  {title} {\bibinfo {title} {Onset of superfluidity in
			$^{4}\mathrm{He}$ films adsorbed on disordered substrates},\ }\href@noop {}
	{\bibfield  {journal} {\bibinfo  {journal} {Phys. Rev. B}\ }\textbf {\bibinfo
			{volume} {55}},\ \bibinfo {pages} {12620} (\bibinfo {year}
		{1997})}\BibitemShut {NoStop}%
	\bibitem [{\citenamefont {Sac{\'e}p{\'e}}\ \emph {et~al.}(2011)\citenamefont
		{Sac{\'e}p{\'e}}, \citenamefont {Dubouchet}, \citenamefont {Chapelier},
		\citenamefont {Sanquer}, \citenamefont {Ovadia}, \citenamefont {Shahar},
		\citenamefont {Feigel'man},\ and\ \citenamefont
		{Ioffe}}]{Sacepe_Ioffe_2011_NatPhys}%
	\BibitemOpen
	\bibfield  {author} {\bibinfo {author} {\bibfnamefont {B.}~\bibnamefont
			{Sac{\'e}p{\'e}}}, \bibinfo {author} {\bibfnamefont {T.}~\bibnamefont
			{Dubouchet}}, \bibinfo {author} {\bibfnamefont {C.}~\bibnamefont
			{Chapelier}}, \bibinfo {author} {\bibfnamefont {M.}~\bibnamefont {Sanquer}},
		\bibinfo {author} {\bibfnamefont {M.}~\bibnamefont {Ovadia}}, \bibinfo
		{author} {\bibfnamefont {D.}~\bibnamefont {Shahar}}, \bibinfo {author}
		{\bibfnamefont {M.}~\bibnamefont {Feigel'man}},\ and\ \bibinfo {author}
		{\bibfnamefont {L.}~\bibnamefont {Ioffe}},\ }\bibfield  {title} {\bibinfo
		{title} {Localization of preformed {Cooper} pairs in disordered
			superconductors},\ }\href@noop {} {\bibfield  {journal} {\bibinfo  {journal}
			{Nat. Phys.}\ }\textbf {\bibinfo {volume} {7}},\ \bibinfo {pages} {239}
		(\bibinfo {year} {2011})}\BibitemShut {NoStop}%
	\bibitem [{\citenamefont {Yu}\ \emph {et~al.}(2012)\citenamefont {Yu},
		\citenamefont {Yin}, \citenamefont {Sullivan}, \citenamefont {Xia},
		\citenamefont {Huan}, \citenamefont {Paduan-Filho}, \citenamefont
		{Oliveira~Jr}, \citenamefont {Haas}, \citenamefont {Steppke}, \citenamefont
		{Miclea}, \citenamefont {Weickert}, \citenamefont {Movshovich}, \citenamefont
		{Mun}, \citenamefont {Scott}, \citenamefont {Zapf},\ and\ \citenamefont
		{Roscilde}}]{Yu_Roscilde_2012_Nature}%
	\BibitemOpen
	\bibfield  {author} {\bibinfo {author} {\bibfnamefont {R.}~\bibnamefont
			{Yu}}, \bibinfo {author} {\bibfnamefont {L.}~\bibnamefont {Yin}}, \bibinfo
		{author} {\bibfnamefont {N.~S.}\ \bibnamefont {Sullivan}}, \bibinfo {author}
		{\bibfnamefont {J.~S.}\ \bibnamefont {Xia}}, \bibinfo {author} {\bibfnamefont
			{C.}~\bibnamefont {Huan}}, \bibinfo {author} {\bibfnamefont {A.}~\bibnamefont
			{Paduan-Filho}}, \bibinfo {author} {\bibfnamefont {N.~F.}\ \bibnamefont
			{Oliveira~Jr}}, \bibinfo {author} {\bibfnamefont {S.}~\bibnamefont {Haas}},
		\bibinfo {author} {\bibfnamefont {A.}~\bibnamefont {Steppke}}, \bibinfo
		{author} {\bibfnamefont {C.~F.}\ \bibnamefont {Miclea}}, \bibinfo {author}
		{\bibfnamefont {F.}~\bibnamefont {Weickert}}, \bibinfo {author}
		{\bibfnamefont {R.}~\bibnamefont {Movshovich}}, \bibinfo {author}
		{\bibfnamefont {E.-D.}\ \bibnamefont {Mun}}, \bibinfo {author} {\bibfnamefont
			{B.~L.}\ \bibnamefont {Scott}}, \bibinfo {author} {\bibfnamefont {V.~S.}\
			\bibnamefont {Zapf}},\ and\ \bibinfo {author} {\bibfnamefont
			{T.}~\bibnamefont {Roscilde}},\ }\bibfield  {title} {\bibinfo {title} {{Bose}
			glass and {Mott} glass of quasiparticles in a doped quantum magnet},\
	}\href@noop {} {\bibfield  {journal} {\bibinfo  {journal} {Nature (London)}\
		}\textbf {\bibinfo {volume} {489}},\ \bibinfo {pages} {379} (\bibinfo {year}
		{2012})}\BibitemShut {NoStop}%
	\bibitem [{\citenamefont {Fallani}\ \emph {et~al.}(2007)\citenamefont
		{Fallani}, \citenamefont {Lye}, \citenamefont {Guarrera}, \citenamefont
		{Fort},\ and\ \citenamefont {Inguscio}}]{Fallani_Inguscio_2007_PRL}%
	\BibitemOpen
	\bibfield  {author} {\bibinfo {author} {\bibfnamefont {L.}~\bibnamefont
			{Fallani}}, \bibinfo {author} {\bibfnamefont {J.~E.}\ \bibnamefont {Lye}},
		\bibinfo {author} {\bibfnamefont {V.}~\bibnamefont {Guarrera}}, \bibinfo
		{author} {\bibfnamefont {C.}~\bibnamefont {Fort}},\ and\ \bibinfo {author}
		{\bibfnamefont {M.}~\bibnamefont {Inguscio}},\ }\bibfield  {title} {\bibinfo
		{title} {Ultracold atoms in a disordered crystal of light: Towards a {Bose}
			glass},\ }\href@noop {} {\bibfield  {journal} {\bibinfo  {journal} {Phys.
				Rev. Lett.}\ }\textbf {\bibinfo {volume} {98}},\ \bibinfo {pages} {130404}
		(\bibinfo {year} {2007})}\BibitemShut {NoStop}%
	\bibitem [{\citenamefont {Deissler}\ \emph {et~al.}(2010)\citenamefont
		{Deissler}, \citenamefont {Zaccanti}, \citenamefont {Roati}, \citenamefont
		{D'Errico}, \citenamefont {Fattori}, \citenamefont {Modugno}, \citenamefont
		{Modugno},\ and\ \citenamefont {Inguscio}}]{Deissler_Inguscio_2010_NatPhys}%
	\BibitemOpen
	\bibfield  {author} {\bibinfo {author} {\bibfnamefont {B.}~\bibnamefont
			{Deissler}}, \bibinfo {author} {\bibfnamefont {M.}~\bibnamefont {Zaccanti}},
		\bibinfo {author} {\bibfnamefont {G.}~\bibnamefont {Roati}}, \bibinfo
		{author} {\bibfnamefont {C.}~\bibnamefont {D'Errico}}, \bibinfo {author}
		{\bibfnamefont {M.}~\bibnamefont {Fattori}}, \bibinfo {author} {\bibfnamefont
			{M.}~\bibnamefont {Modugno}}, \bibinfo {author} {\bibfnamefont
			{G.}~\bibnamefont {Modugno}},\ and\ \bibinfo {author} {\bibfnamefont
			{M.}~\bibnamefont {Inguscio}},\ }\bibfield  {title} {\bibinfo {title}
		{Delocalization of a disordered bosonic system by repulsive interactions},\
	}\href@noop {} {\bibfield  {journal} {\bibinfo  {journal} {Nat. Phys.}\
		}\textbf {\bibinfo {volume} {6}},\ \bibinfo {pages} {354} (\bibinfo {year}
		{2010})}\BibitemShut {NoStop}%
	\bibitem [{\citenamefont {Pasienski}\ \emph {et~al.}(2010)\citenamefont
		{Pasienski}, \citenamefont {McKay}, \citenamefont {White},\ and\
		\citenamefont {DeMarco}}]{Pasienski_DeMarco_2010_NatPhys}%
	\BibitemOpen
	\bibfield  {author} {\bibinfo {author} {\bibfnamefont {M.}~\bibnamefont
			{Pasienski}}, \bibinfo {author} {\bibfnamefont {D.}~\bibnamefont {McKay}},
		\bibinfo {author} {\bibfnamefont {M.}~\bibnamefont {White}},\ and\ \bibinfo
		{author} {\bibfnamefont {B.}~\bibnamefont {DeMarco}},\ }\bibfield  {title}
	{\bibinfo {title} {A disordered insulator in an optical lattice},\
	}\href@noop {} {\bibfield  {journal} {\bibinfo  {journal} {Nat. Phys.}\
		}\textbf {\bibinfo {volume} {6}},\ \bibinfo {pages} {677} (\bibinfo {year}
		{2010})}\BibitemShut {NoStop}%
	\bibitem [{\citenamefont {Gadway}\ \emph {et~al.}(2011)\citenamefont {Gadway},
		\citenamefont {Pertot}, \citenamefont {Reeves}, \citenamefont {Vogt},\ and\
		\citenamefont {Schneble}}]{Gadway_Schneble_2011_PRL}%
	\BibitemOpen
	\bibfield  {author} {\bibinfo {author} {\bibfnamefont {B.}~\bibnamefont
			{Gadway}}, \bibinfo {author} {\bibfnamefont {D.}~\bibnamefont {Pertot}},
		\bibinfo {author} {\bibfnamefont {J.}~\bibnamefont {Reeves}}, \bibinfo
		{author} {\bibfnamefont {M.}~\bibnamefont {Vogt}},\ and\ \bibinfo {author}
		{\bibfnamefont {D.}~\bibnamefont {Schneble}},\ }\bibfield  {title} {\bibinfo
		{title} {Glassy behavior in a binary atomic mixture},\ }\href@noop {}
	{\bibfield  {journal} {\bibinfo  {journal} {Phys. Rev. Lett.}\ }\textbf
		{\bibinfo {volume} {107}},\ \bibinfo {pages} {145306} (\bibinfo {year}
		{2011})}\BibitemShut {NoStop}%
	\bibitem [{\citenamefont {D'Errico}\ \emph {et~al.}(2014)\citenamefont
		{D'Errico}, \citenamefont {Lucioni}, \citenamefont {Tanzi}, \citenamefont
		{Gori}, \citenamefont {Roux}, \citenamefont {McCulloch}, \citenamefont
		{Giamarchi}, \citenamefont {Inguscio},\ and\ \citenamefont
		{Modugno}}]{D'Errico_Modugno_2014_PRL}%
	\BibitemOpen
	\bibfield  {author} {\bibinfo {author} {\bibfnamefont {C.}~\bibnamefont
			{D'Errico}}, \bibinfo {author} {\bibfnamefont {E.}~\bibnamefont {Lucioni}},
		\bibinfo {author} {\bibfnamefont {L.}~\bibnamefont {Tanzi}}, \bibinfo
		{author} {\bibfnamefont {L.}~\bibnamefont {Gori}}, \bibinfo {author}
		{\bibfnamefont {G.}~\bibnamefont {Roux}}, \bibinfo {author} {\bibfnamefont
			{I.~P.}\ \bibnamefont {McCulloch}}, \bibinfo {author} {\bibfnamefont
			{T.}~\bibnamefont {Giamarchi}}, \bibinfo {author} {\bibfnamefont
			{M.}~\bibnamefont {Inguscio}},\ and\ \bibinfo {author} {\bibfnamefont
			{G.}~\bibnamefont {Modugno}},\ }\bibfield  {title} {\bibinfo {title}
		{Observation of a disordered bosonic insulator from weak to strong
			interactions},\ }\href@noop {} {\bibfield  {journal} {\bibinfo  {journal}
			{Phys. Rev. Lett.}\ }\textbf {\bibinfo {volume} {113}},\ \bibinfo {pages}
		{095301} (\bibinfo {year} {2014})}\BibitemShut {NoStop}%
	\bibitem [{\citenamefont {Meldgin}\ \emph {et~al.}(2016)\citenamefont
		{Meldgin}, \citenamefont {Ray}, \citenamefont {Russ}, \citenamefont {Chen},
		\citenamefont {Ceperley},\ and\ \citenamefont
		{DeMarco}}]{Meldgin_DeMarco_2016_NatPhys}%
	\BibitemOpen
	\bibfield  {author} {\bibinfo {author} {\bibfnamefont {C.}~\bibnamefont
			{Meldgin}}, \bibinfo {author} {\bibfnamefont {U.}~\bibnamefont {Ray}},
		\bibinfo {author} {\bibfnamefont {P.}~\bibnamefont {Russ}}, \bibinfo {author}
		{\bibfnamefont {D.}~\bibnamefont {Chen}}, \bibinfo {author} {\bibfnamefont
			{D.~M.}\ \bibnamefont {Ceperley}},\ and\ \bibinfo {author} {\bibfnamefont
			{B.}~\bibnamefont {DeMarco}},\ }\bibfield  {title} {\bibinfo {title} {Probing
			the {Bose} glass--superfluid transition using quantum quenches of disorder},\
	}\href@noop {} {\bibfield  {journal} {\bibinfo  {journal} {Nat. Phys.}\
		}\textbf {\bibinfo {volume} {12}},\ \bibinfo {pages} {646} (\bibinfo {year}
		{2016})}\BibitemShut {NoStop}%
	\bibitem [{\citenamefont {Rapsch}\ \emph {et~al.}(1999)\citenamefont {Rapsch},
		\citenamefont {Schollw\"ock},\ and\ \citenamefont
		{Zwerger}}]{Rapsch_Zwerger_1999_EPL}%
	\BibitemOpen
	\bibfield  {author} {\bibinfo {author} {\bibfnamefont {S.}~\bibnamefont
			{Rapsch}}, \bibinfo {author} {\bibfnamefont {U.}~\bibnamefont
			{Schollw\"ock}},\ and\ \bibinfo {author} {\bibfnamefont {W.}~\bibnamefont
			{Zwerger}},\ }\bibfield  {title} {\bibinfo {title} {Density matrix
			renormalization group for disordered bosons in one dimension},\ }\href@noop
	{} {\bibfield  {journal} {\bibinfo  {journal} {Europhys. Lett.}\ }\textbf
		{\bibinfo {volume} {46}},\ \bibinfo {pages} {559} (\bibinfo {year}
		{1999})}\BibitemShut {NoStop}%
	\bibitem [{\citenamefont {Lugan}\ \emph
		{et~al.}(2007{\natexlab{a}})\citenamefont {Lugan}, \citenamefont {Cl\'ement},
		\citenamefont {Bouyer}, \citenamefont {Aspect}, \citenamefont {Lewenstein},\
		and\ \citenamefont {Sanchez-Palencia}}]{Lugan_PRL_2007_1}%
	\BibitemOpen
	\bibfield  {author} {\bibinfo {author} {\bibfnamefont {P.}~\bibnamefont
			{Lugan}}, \bibinfo {author} {\bibfnamefont {D.}~\bibnamefont {Cl\'ement}},
		\bibinfo {author} {\bibfnamefont {P.}~\bibnamefont {Bouyer}}, \bibinfo
		{author} {\bibfnamefont {A.}~\bibnamefont {Aspect}}, \bibinfo {author}
		{\bibfnamefont {M.}~\bibnamefont {Lewenstein}},\ and\ \bibinfo {author}
		{\bibfnamefont {L.}~\bibnamefont {Sanchez-Palencia}},\ }\bibfield  {title}
	{\bibinfo {title} {Ultracold {Bose} gases in {1D} disorder: From {Lifshits}
			glass to {Bose}-{Einstein} condensate},\ }\href@noop {} {\bibfield  {journal}
		{\bibinfo  {journal} {Phys. Rev. Lett.}\ }\textbf {\bibinfo {volume} {98}},\
		\bibinfo {pages} {170403} (\bibinfo {year} {2007}{\natexlab{a}})}\BibitemShut
	{NoStop}%
	\bibitem [{\citenamefont {Lugan}\ \emph
		{et~al.}(2007{\natexlab{b}})\citenamefont {Lugan}, \citenamefont {Cl\'ement},
		\citenamefont {Bouyer}, \citenamefont {Aspect},\ and\ \citenamefont
		{Sanchez-Palencia}}]{Lugan_PRL_2007_2}%
	\BibitemOpen
	\bibfield  {author} {\bibinfo {author} {\bibfnamefont {P.}~\bibnamefont
			{Lugan}}, \bibinfo {author} {\bibfnamefont {D.}~\bibnamefont {Cl\'ement}},
		\bibinfo {author} {\bibfnamefont {P.}~\bibnamefont {Bouyer}}, \bibinfo
		{author} {\bibfnamefont {A.}~\bibnamefont {Aspect}},\ and\ \bibinfo {author}
		{\bibfnamefont {L.}~\bibnamefont {Sanchez-Palencia}},\ }\bibfield  {title}
	{\bibinfo {title} {Anderson localization of {Bogolyubov} quasiparticles in
			interacting {Bose}-{Einstein} condensates},\ }\href@noop {} {\bibfield
		{journal} {\bibinfo  {journal} {Phys. Rev. Lett.}\ }\textbf {\bibinfo
			{volume} {99}},\ \bibinfo {pages} {180402} (\bibinfo {year}
		{2007}{\natexlab{b}})}\BibitemShut {NoStop}%
	\bibitem [{\citenamefont {Roux}\ \emph {et~al.}(2008)\citenamefont {Roux},
		\citenamefont {Barthel}, \citenamefont {McCulloch}, \citenamefont {Kollath},
		\citenamefont {Schollw\"ock},\ and\ \citenamefont
		{Giamarchi}}]{Roux_Giamarchi_2008_PRA}%
	\BibitemOpen
	\bibfield  {author} {\bibinfo {author} {\bibfnamefont {G.}~\bibnamefont
			{Roux}}, \bibinfo {author} {\bibfnamefont {T.}~\bibnamefont {Barthel}},
		\bibinfo {author} {\bibfnamefont {I.~P.}\ \bibnamefont {McCulloch}}, \bibinfo
		{author} {\bibfnamefont {C.}~\bibnamefont {Kollath}}, \bibinfo {author}
		{\bibfnamefont {U.}~\bibnamefont {Schollw\"ock}},\ and\ \bibinfo {author}
		{\bibfnamefont {T.}~\bibnamefont {Giamarchi}},\ }\bibfield  {title} {\bibinfo
		{title} {Quasiperiodic {Bose}-{Hubbard} model and localization in
			one-dimensional cold atomic gases},\ }\href@noop {} {\bibfield  {journal}
		{\bibinfo  {journal} {Phys. Rev. A}\ }\textbf {\bibinfo {volume} {78}},\
		\bibinfo {pages} {023628} (\bibinfo {year} {2008})}\BibitemShut {NoStop}%
	\bibitem [{\citenamefont {Bissbort}\ and\ \citenamefont
		{Hofstetter}(2009)}]{Bissbort_Hofstetter_2009_EPL}%
	\BibitemOpen
	\bibfield  {author} {\bibinfo {author} {\bibfnamefont {U.}~\bibnamefont
			{Bissbort}}\ and\ \bibinfo {author} {\bibfnamefont {W.}~\bibnamefont
			{Hofstetter}},\ }\bibfield  {title} {\bibinfo {title} {Stochastic mean-field
			theory for the disordered {Bose}-{Hubbard} model},\ }\href@noop {} {\bibfield
		{journal} {\bibinfo  {journal} {Europhys. Lett.}\ }\textbf {\bibinfo
			{volume} {86}},\ \bibinfo {pages} {50007} (\bibinfo {year}
		{2009})}\BibitemShut {NoStop}%
	\bibitem [{\citenamefont {S\"oyler}\ \emph {et~al.}(2011)\citenamefont
		{S\"oyler}, \citenamefont {Kiselev}, \citenamefont {Prokof'ev},\ and\
		\citenamefont {Svistunov}}]{Soeyler_Svistunov_2011_PRL}%
	\BibitemOpen
	\bibfield  {author} {\bibinfo {author} {\bibfnamefont {{\c S}.~G.}\
			\bibnamefont {S\"oyler}}, \bibinfo {author} {\bibfnamefont {M.}~\bibnamefont
			{Kiselev}}, \bibinfo {author} {\bibfnamefont {N.~V.}\ \bibnamefont
			{Prokof'ev}},\ and\ \bibinfo {author} {\bibfnamefont {B.~V.}\ \bibnamefont
			{Svistunov}},\ }\bibfield  {title} {\bibinfo {title} {Phase diagram of the
			commensurate two-dimensional disordered {Bose}-{Hubbard} model},\ }\href@noop
	{} {\bibfield  {journal} {\bibinfo  {journal} {Phys. Rev. Lett.}\ }\textbf
		{\bibinfo {volume} {107}},\ \bibinfo {pages} {185301} (\bibinfo {year}
		{2011})}\BibitemShut {NoStop}%
	\bibitem [{\citenamefont {Carleo}\ \emph {et~al.}(2013)\citenamefont {Carleo},
		\citenamefont {Bo\'eris}, \citenamefont {Holzmann},\ and\ \citenamefont
		{Sanchez-Palencia}}]{Carleo_PRL_2013}%
	\BibitemOpen
	\bibfield  {author} {\bibinfo {author} {\bibfnamefont {G.}~\bibnamefont
			{Carleo}}, \bibinfo {author} {\bibfnamefont {G.}~\bibnamefont {Bo\'eris}},
		\bibinfo {author} {\bibfnamefont {M.}~\bibnamefont {Holzmann}},\ and\
		\bibinfo {author} {\bibfnamefont {L.}~\bibnamefont {Sanchez-Palencia}},\
	}\bibfield  {title} {\bibinfo {title} {Universal superfluid transition and
			transport properties of two-dimensional dirty bosons},\ }\href@noop {}
	{\bibfield  {journal} {\bibinfo  {journal} {Phys. Rev. Lett.}\ }\textbf
		{\bibinfo {volume} {111}},\ \bibinfo {pages} {050406} (\bibinfo {year}
		{2013})}\BibitemShut {NoStop}%
	\bibitem [{\citenamefont {Niederle}\ and\ \citenamefont
		{Rieger}(2015)}]{Niederle_Rieger_2015_PRA}%
	\BibitemOpen
	\bibfield  {author} {\bibinfo {author} {\bibfnamefont {A.~E.}\ \bibnamefont
			{Niederle}}\ and\ \bibinfo {author} {\bibfnamefont {H.}~\bibnamefont
			{Rieger}},\ }\bibfield  {title} {\bibinfo {title} {Bosons in a
			two-dimensional bichromatic quasiperiodic potential: Analysis of the disorder
			in the {Bose}-{Hubbard} parameters and phase diagrams},\ }\href@noop {}
	{\bibfield  {journal} {\bibinfo  {journal} {Phys. Rev. A}\ }\textbf {\bibinfo
			{volume} {91}},\ \bibinfo {pages} {043632} (\bibinfo {year}
		{2015})}\BibitemShut {NoStop}%
	\bibitem [{\citenamefont {Zhang}\ \emph {et~al.}(2015)\citenamefont {Zhang},
		\citenamefont {Safavi-Naini},\ and\ \citenamefont
		{Capogrosso-Sansone}}]{Zhang_Capogrosso-Sansone_2015_PRA}%
	\BibitemOpen
	\bibfield  {author} {\bibinfo {author} {\bibfnamefont {C.}~\bibnamefont
			{Zhang}}, \bibinfo {author} {\bibfnamefont {A.}~\bibnamefont
			{Safavi-Naini}},\ and\ \bibinfo {author} {\bibfnamefont {B.}~\bibnamefont
			{Capogrosso-Sansone}},\ }\bibfield  {title} {\bibinfo {title} {Equilibrium
			phases of two-dimensional bosons in quasiperiodic lattices},\ }\href@noop {}
	{\bibfield  {journal} {\bibinfo  {journal} {Phys. Rev. A}\ }\textbf {\bibinfo
			{volume} {91}},\ \bibinfo {pages} {031604(R)} (\bibinfo {year}
		{2015})}\BibitemShut {NoStop}%
	\bibitem [{\citenamefont {Gerster}\ \emph {et~al.}(2016)\citenamefont
		{Gerster}, \citenamefont {Rizzi}, \citenamefont {Tschirsich}, \citenamefont
		{Silvi}, \citenamefont {Fazio},\ and\ \citenamefont
		{Montangero}}]{Gerster_Montangero_2016_NJP}%
	\BibitemOpen
	\bibfield  {author} {\bibinfo {author} {\bibfnamefont {M.}~\bibnamefont
			{Gerster}}, \bibinfo {author} {\bibfnamefont {M.}~\bibnamefont {Rizzi}},
		\bibinfo {author} {\bibfnamefont {F.}~\bibnamefont {Tschirsich}}, \bibinfo
		{author} {\bibfnamefont {P.}~\bibnamefont {Silvi}}, \bibinfo {author}
		{\bibfnamefont {R.}~\bibnamefont {Fazio}},\ and\ \bibinfo {author}
		{\bibfnamefont {S.}~\bibnamefont {Montangero}},\ }\bibfield  {title}
	{\bibinfo {title} {Superfluid density and quasi-long-range order in the
			one-dimensional disordered {Bose}-{Hubbard} model},\ }\href@noop {}
	{\bibfield  {journal} {\bibinfo  {journal} {New J. Phys.}\ }\textbf {\bibinfo
			{volume} {18}},\ \bibinfo {pages} {015015} (\bibinfo {year}
		{2016})}\BibitemShut {NoStop}%
	\bibitem [{\citenamefont {Yao}\ \emph {et~al.}(2020)\citenamefont {Yao},
		\citenamefont {Giamarchi},\ and\ \citenamefont
		{Sanchez-Palencia}}]{Yao_Sanchez-Palencia_2020_PRL}%
	\BibitemOpen
	\bibfield  {author} {\bibinfo {author} {\bibfnamefont {H.}~\bibnamefont
			{Yao}}, \bibinfo {author} {\bibfnamefont {T.}~\bibnamefont {Giamarchi}},\
		and\ \bibinfo {author} {\bibfnamefont {L.}~\bibnamefont {Sanchez-Palencia}},\
	}\bibfield  {title} {\bibinfo {title} {Lieb-{Liniger} bosons in a shallow
			quasiperiodic potential: {Bose} glass phase and fractal {Mott} lobes},\
	}\href@noop {} {\bibfield  {journal} {\bibinfo  {journal} {Phys. Rev. Lett.}\
		}\textbf {\bibinfo {volume} {125}},\ \bibinfo {pages} {060401} (\bibinfo
		{year} {2020})}\BibitemShut {NoStop}%
	\bibitem [{\citenamefont {Johnstone}\ \emph {et~al.}(2021)\citenamefont
		{Johnstone}, \citenamefont {Ohberg},\ and\ \citenamefont
		{Duncan}}]{Johnstone_Duncan_2021_JPA}%
	\BibitemOpen
	\bibfield  {author} {\bibinfo {author} {\bibfnamefont {D.}~\bibnamefont
			{Johnstone}}, \bibinfo {author} {\bibfnamefont {P.}~\bibnamefont {Ohberg}},\
		and\ \bibinfo {author} {\bibfnamefont {C.~W.}\ \bibnamefont {Duncan}},\
	}\bibfield  {title} {\bibinfo {title} {The mean-field {Bose} glass in
			quasicrystalline systems},\ }\href@noop {} {\bibfield  {journal} {\bibinfo
			{journal} {J. Phys. A}\ }\textbf {\bibinfo {volume} {54}},\ \bibinfo {pages}
		{395001} (\bibinfo {year} {2021})}\BibitemShut {NoStop}%
	\bibitem [{\citenamefont {Gautier}\ \emph {et~al.}(2021)\citenamefont
		{Gautier}, \citenamefont {Yao},\ and\ \citenamefont
		{Sanchez-Palencia}}]{Gautier_PRL_2021}%
	\BibitemOpen
	\bibfield  {author} {\bibinfo {author} {\bibfnamefont {R.}~\bibnamefont
			{Gautier}}, \bibinfo {author} {\bibfnamefont {H.}~\bibnamefont {Yao}},\ and\
		\bibinfo {author} {\bibfnamefont {L.}~\bibnamefont {Sanchez-Palencia}},\
	}\bibfield  {title} {\bibinfo {title} {Strongly interacting bosons in a
			two-dimensional quasicrystal lattice},\ }\href@noop {} {\bibfield  {journal}
		{\bibinfo  {journal} {Phys. Rev. Lett.}\ }\textbf {\bibinfo {volume} {126}},\
		\bibinfo {pages} {110401} (\bibinfo {year} {2021})}\BibitemShut {NoStop}%
	\bibitem [{\citenamefont {Zhu}\ \emph {et~al.}(2023)\citenamefont {Zhu},
		\citenamefont {Yao},\ and\ \citenamefont {Sanchez-Palencia}}]{Zhu_PRL_2023}%
	\BibitemOpen
	\bibfield  {author} {\bibinfo {author} {\bibfnamefont {Z.}~\bibnamefont
			{Zhu}}, \bibinfo {author} {\bibfnamefont {H.}~\bibnamefont {Yao}},\ and\
		\bibinfo {author} {\bibfnamefont {L.}~\bibnamefont {Sanchez-Palencia}},\
	}\bibfield  {title} {\bibinfo {title} {Thermodynamic phase diagram of
			two-dimensional bosons in a quasicrystal potential},\ }\href@noop {}
	{\bibfield  {journal} {\bibinfo  {journal} {Phys. Rev. Lett.}\ }\textbf
		{\bibinfo {volume} {130}},\ \bibinfo {pages} {220402} (\bibinfo {year}
		{2023})}\BibitemShut {NoStop}%
	\bibitem [{\citenamefont {Zhu}\ \emph {et~al.}(2024)\citenamefont {Zhu},
		\citenamefont {Yu}, \citenamefont {Johnstone},\ and\ \citenamefont
		{Sanchez-Palencia}}]{Zhu_PRA_2024}%
	\BibitemOpen
	\bibfield  {author} {\bibinfo {author} {\bibfnamefont {Z.}~\bibnamefont
			{Zhu}}, \bibinfo {author} {\bibfnamefont {S.}~\bibnamefont {Yu}}, \bibinfo
		{author} {\bibfnamefont {D.}~\bibnamefont {Johnstone}},\ and\ \bibinfo
		{author} {\bibfnamefont {L.}~\bibnamefont {Sanchez-Palencia}},\ }\bibfield
	{title} {\bibinfo {title} {Localization and spectral structure in
			two-dimensional quasicrystal potentials},\ }\href@noop {} {\bibfield
		{journal} {\bibinfo  {journal} {Phys. Rev. A}\ }\textbf {\bibinfo {volume}
			{109}},\ \bibinfo {pages} {013314} (\bibinfo {year} {2024})}\BibitemShut
	{NoStop}%
	\bibitem [{\citenamefont {Molignini}\ and\ \citenamefont
		{Chakrabarti}(2025)}]{Molignini_PRR_2025_2}%
	\BibitemOpen
	\bibfield  {author} {\bibinfo {author} {\bibfnamefont {P.}~\bibnamefont
			{Molignini}}\ and\ \bibinfo {author} {\bibfnamefont {B.}~\bibnamefont
			{Chakrabarti}},\ }\bibfield  {title} {\bibinfo {title} {Stability of dipolar
			bosons in a quasiperiodic potential},\ }\href@noop {} {\bibfield  {journal}
		{\bibinfo  {journal} {Phys. Rev. Res.}\ }\textbf {\bibinfo {volume} {7}},\
		\bibinfo {pages} {023237} (\bibinfo {year} {2025})}\BibitemShut {NoStop}%
	\bibitem [{\citenamefont {Zeng}\ \emph {et~al.}(2025)\citenamefont {Zeng},
		\citenamefont {Zhu},\ and\ \citenamefont {He}}]{Zeng_PRA_2025}%
	\BibitemOpen
	\bibfield  {author} {\bibinfo {author} {\bibfnamefont {J.-H.}\ \bibnamefont
			{Zeng}}, \bibinfo {author} {\bibfnamefont {Q.}~\bibnamefont {Zhu}},\ and\
		\bibinfo {author} {\bibfnamefont {L.}~\bibnamefont {He}},\ }\bibfield
	{title} {\bibinfo {title} {Interaction-induced moir\'e systems in twisted
			bilayer optical lattices},\ }\href@noop {} {\bibfield  {journal} {\bibinfo
			{journal} {Phys. Rev. A}\ }\textbf {\bibinfo {volume} {111}},\ \bibinfo
		{pages} {063317} (\bibinfo {year} {2025})}\BibitemShut {NoStop}%
	\bibitem [{\citenamefont {Meng}\ \emph {et~al.}(2023)\citenamefont {Meng},
		\citenamefont {Wang}, \citenamefont {Han}, \citenamefont {Liu}, \citenamefont
		{Wen}, \citenamefont {Gao}, \citenamefont {Wang}, \citenamefont {Chin},\ and\
		\citenamefont {Zhang}}]{Meng_2023_Nature}%
	\BibitemOpen
	\bibfield  {author} {\bibinfo {author} {\bibfnamefont {Z.}~\bibnamefont
			{Meng}}, \bibinfo {author} {\bibfnamefont {L.}~\bibnamefont {Wang}}, \bibinfo
		{author} {\bibfnamefont {W.}~\bibnamefont {Han}}, \bibinfo {author}
		{\bibfnamefont {F.}~\bibnamefont {Liu}}, \bibinfo {author} {\bibfnamefont
			{K.}~\bibnamefont {Wen}}, \bibinfo {author} {\bibfnamefont {C.}~\bibnamefont
			{Gao}}, \bibinfo {author} {\bibfnamefont {P.}~\bibnamefont {Wang}}, \bibinfo
		{author} {\bibfnamefont {C.}~\bibnamefont {Chin}},\ and\ \bibinfo {author}
		{\bibfnamefont {J.}~\bibnamefont {Zhang}},\ }\bibfield  {title} {\bibinfo
		{title} {Atomic {Bose--Einstein} condensate in twisted-bilayer optical
			lattices},\ }\href@noop {} {\bibfield  {journal} {\bibinfo  {journal} {Nature
				(London)}\ }\textbf {\bibinfo {volume} {615}},\ \bibinfo {pages} {231}
		(\bibinfo {year} {2023})}\BibitemShut {NoStop}%
	\bibitem [{\citenamefont {Johnstone}\ \emph {et~al.}(2024)\citenamefont
		{Johnstone}, \citenamefont {Mishra}, \citenamefont {Zhu}, \citenamefont
		{Yao},\ and\ \citenamefont {Sanchez-Palencia}}]{Johnstone_PRR_2024}%
	\BibitemOpen
	\bibfield  {author} {\bibinfo {author} {\bibfnamefont {D.}~\bibnamefont
			{Johnstone}}, \bibinfo {author} {\bibfnamefont {S.}~\bibnamefont {Mishra}},
		\bibinfo {author} {\bibfnamefont {Z.}~\bibnamefont {Zhu}}, \bibinfo {author}
		{\bibfnamefont {H.}~\bibnamefont {Yao}},\ and\ \bibinfo {author}
		{\bibfnamefont {L.}~\bibnamefont {Sanchez-Palencia}},\ }\bibfield  {title}
	{\bibinfo {title} {Weak superfluidity in twisted optical potentials},\
	}\href@noop {} {\bibfield  {journal} {\bibinfo  {journal} {Phys. Rev. Res.}\
		}\textbf {\bibinfo {volume} {6}},\ \bibinfo {pages} {L042066} (\bibinfo
		{year} {2024})}\BibitemShut {NoStop}%
	\bibitem [{\citenamefont {Johnstone}\ \emph {et~al.}(2025)\citenamefont
		{Johnstone}, \citenamefont {Mishra}, \citenamefont {Zhu},\ and\ \citenamefont
		{Sanchez-Palencia}}]{Johnstone_PRA_2025}%
	\BibitemOpen
	\bibfield  {author} {\bibinfo {author} {\bibfnamefont {D.}~\bibnamefont
			{Johnstone}}, \bibinfo {author} {\bibfnamefont {S.}~\bibnamefont {Mishra}},
		\bibinfo {author} {\bibfnamefont {Z.}~\bibnamefont {Zhu}},\ and\ \bibinfo
		{author} {\bibfnamefont {L.}~\bibnamefont {Sanchez-Palencia}},\ }\bibfield
	{title} {\bibinfo {title} {Effective tight-binding models in optical moir\'e
			potentials},\ }\href@noop {} {\bibfield  {journal} {\bibinfo  {journal}
			{Phys. Rev. A}\ }\textbf {\bibinfo {volume} {111}},\ \bibinfo {pages}
		{043305} (\bibinfo {year} {2025})}\BibitemShut {NoStop}%
	\bibitem [{\citenamefont {Yu}\ \emph {et~al.}(2024)\citenamefont {Yu},
		\citenamefont {Bhave}, \citenamefont {Reeve}, \citenamefont {Song},\ and\
		\citenamefont {Schneider}}]{Yu_Schneider_2024_Nature}%
	\BibitemOpen
	\bibfield  {author} {\bibinfo {author} {\bibfnamefont {J.-C.}\ \bibnamefont
			{Yu}}, \bibinfo {author} {\bibfnamefont {S.}~\bibnamefont {Bhave}}, \bibinfo
		{author} {\bibfnamefont {L.}~\bibnamefont {Reeve}}, \bibinfo {author}
		{\bibfnamefont {B.}~\bibnamefont {Song}},\ and\ \bibinfo {author}
		{\bibfnamefont {U.}~\bibnamefont {Schneider}},\ }\bibfield  {title} {\bibinfo
		{title} {Observing the two-dimensional {Bose} glass in an optical
			quasicrystal},\ }\href@noop {} {\bibfield  {journal} {\bibinfo  {journal}
			{Nature (London)}\ }\textbf {\bibinfo {volume} {633}},\ \bibinfo {pages}
		{338} (\bibinfo {year} {2024})}\BibitemShut {NoStop}%
	\bibitem [{\citenamefont {Wan}\ \emph {et~al.}()\citenamefont {Wan},
		\citenamefont {Gao},\ and\ \citenamefont {Shi}}]{Wan_2024_arXiv}%
	\BibitemOpen
	\bibfield  {author} {\bibinfo {author} {\bibfnamefont {X.-T.}\ \bibnamefont
			{Wan}}, \bibinfo {author} {\bibfnamefont {C.}~\bibnamefont {Gao}},\ and\
		\bibinfo {author} {\bibfnamefont {Z.-Y.}\ \bibnamefont {Shi}},\ }\bibfield
	{title} {\bibinfo {title} {Fractal spectrum in twisted bilayer optical
			lattice},\ }\href@noop {} {\bibinfo  {journal} {arXiv:2404.08211}\
	}\BibitemShut {NoStop}%
	\bibitem [{\citenamefont {Krauth}\ \emph {et~al.}(1992)\citenamefont {Krauth},
		\citenamefont {Caffarel},\ and\ \citenamefont {Bouchaud}}]{Krauth_PRB_1992}%
	\BibitemOpen
	\bibfield  {journal} {  }\bibfield  {author} {\bibinfo {author} {\bibfnamefont
			{W.}~\bibnamefont {Krauth}}, \bibinfo {author} {\bibfnamefont
			{M.}~\bibnamefont {Caffarel}},\ and\ \bibinfo {author} {\bibfnamefont
			{J.-P.}\ \bibnamefont {Bouchaud}},\ }\bibfield  {title} {\bibinfo {title}
		{{Gutzwiller} wave function for a model of strongly interacting bosons},\
	}\href@noop {} {\bibfield  {journal} {\bibinfo  {journal} {Phys. Rev. B}\
		}\textbf {\bibinfo {volume} {45}},\ \bibinfo {pages} {3137} (\bibinfo {year}
		{1992})}\BibitemShut {NoStop}%
	\bibitem [{\citenamefont {Jaksch}\ \emph {et~al.}(1998)\citenamefont {Jaksch},
		\citenamefont {Bruder}, \citenamefont {Cirac}, \citenamefont {Gardiner},\
		and\ \citenamefont {Zoller}}]{Jaksch_1998_PRL}%
	\BibitemOpen
	\bibfield  {author} {\bibinfo {author} {\bibfnamefont {D.}~\bibnamefont
			{Jaksch}}, \bibinfo {author} {\bibfnamefont {C.}~\bibnamefont {Bruder}},
		\bibinfo {author} {\bibfnamefont {J.~I.}\ \bibnamefont {Cirac}}, \bibinfo
		{author} {\bibfnamefont {C.~W.}\ \bibnamefont {Gardiner}},\ and\ \bibinfo
		{author} {\bibfnamefont {P.}~\bibnamefont {Zoller}},\ }\bibfield  {title}
	{\bibinfo {title} {Cold bosonic atoms in optical lattices},\ }\href@noop {}
	{\bibfield  {journal} {\bibinfo  {journal} {Phys. Rev. Lett.}\ }\textbf
		{\bibinfo {volume} {81}},\ \bibinfo {pages} {3108} (\bibinfo {year}
		{1998})}\BibitemShut {NoStop}%
	\bibitem [{\citenamefont {Lanat{\`a}}\ \emph {et~al.}(2012)\citenamefont
		{Lanat{\`a}}, \citenamefont {Strand}, \citenamefont {Dai},\ and\
		\citenamefont {Hellsing}}]{Lanata_PRB_2012}%
	\BibitemOpen
	\bibfield  {author} {\bibinfo {author} {\bibfnamefont {N.}~\bibnamefont
			{Lanat{\`a}}}, \bibinfo {author} {\bibfnamefont {H.~U.~R.}\ \bibnamefont
			{Strand}}, \bibinfo {author} {\bibfnamefont {X.}~\bibnamefont {Dai}},\ and\
		\bibinfo {author} {\bibfnamefont {B.}~\bibnamefont {Hellsing}},\ }\bibfield
	{title} {\bibinfo {title} {Efficient implementation of the {Gutzwiller}
			variational method},\ }\href@noop {} {\bibfield  {journal} {\bibinfo
			{journal} {Phys. Rev. B}\ }\textbf {\bibinfo {volume} {85}},\ \bibinfo
		{pages} {035133} (\bibinfo {year} {2012})}\BibitemShut {NoStop}%
	\bibitem [{\citenamefont {Sheshadri}\ \emph {et~al.}(1995)\citenamefont
		{Sheshadri}, \citenamefont {Krishnamurthy}, \citenamefont {Pandit},\ and\
		\citenamefont {Ramakrishnan}}]{Sheshadri_1995_PRL}%
	\BibitemOpen
	\bibfield  {author} {\bibinfo {author} {\bibfnamefont {K.}~\bibnamefont
			{Sheshadri}}, \bibinfo {author} {\bibfnamefont {H.~R.}\ \bibnamefont
			{Krishnamurthy}}, \bibinfo {author} {\bibfnamefont {R.}~\bibnamefont
			{Pandit}},\ and\ \bibinfo {author} {\bibfnamefont {T.~V.}\ \bibnamefont
			{Ramakrishnan}},\ }\bibfield  {title} {\bibinfo {title} {Percolation-enhanced
			localization in the disordered bosonic {Hubbard} model},\ }\href@noop {}
	{\bibfield  {journal} {\bibinfo  {journal} {Phys. Rev. Lett.}\ }\textbf
		{\bibinfo {volume} {75}},\ \bibinfo {pages} {4075} (\bibinfo {year}
		{1995})}\BibitemShut {NoStop}%
	\bibitem [{\citenamefont {Buonsante}\ \emph {et~al.}(2009)\citenamefont
		{Buonsante}, \citenamefont {Massel}, \citenamefont {Penna},\ and\
		\citenamefont {Vezzani}}]{Buonsante_2009_PRA}%
	\BibitemOpen
	\bibfield  {author} {\bibinfo {author} {\bibfnamefont {P.}~\bibnamefont
			{Buonsante}}, \bibinfo {author} {\bibfnamefont {F.}~\bibnamefont {Massel}},
		\bibinfo {author} {\bibfnamefont {V.}~\bibnamefont {Penna}},\ and\ \bibinfo
		{author} {\bibfnamefont {A.}~\bibnamefont {Vezzani}},\ }\bibfield  {title}
	{\bibinfo {title} {{Gutzwiller} approach to the {Bose}-{Hubbard} model with
			random local impurities},\ }\href@noop {} {\bibfield  {journal} {\bibinfo
			{journal} {Phys. Rev. A}\ }\textbf {\bibinfo {volume} {79}},\ \bibinfo
		{pages} {013623} (\bibinfo {year} {2009})}\BibitemShut {NoStop}%
	\bibitem [{\citenamefont {Kemburi}\ and\ \citenamefont
		{Scarola}(2012)}]{Kemburi_2012_PRB}%
	\BibitemOpen
	\bibfield  {author} {\bibinfo {author} {\bibfnamefont {B.~M.}\ \bibnamefont
			{Kemburi}}\ and\ \bibinfo {author} {\bibfnamefont {V.~W.}\ \bibnamefont
			{Scarola}},\ }\bibfield  {title} {\bibinfo {title} {Percolation-enhanced
			supersolids in the extended {Bose}-{Hubbard} model},\ }\href@noop {}
	{\bibfield  {journal} {\bibinfo  {journal} {Phys. Rev. B}\ }\textbf {\bibinfo
			{volume} {85}},\ \bibinfo {pages} {020501(R)} (\bibinfo {year}
		{2012})}\BibitemShut {NoStop}%
	\bibitem [{\citenamefont {Niederle}\ and\ \citenamefont
		{Rieger}(2013)}]{Niederle_2013_njp}%
	\BibitemOpen
	\bibfield  {author} {\bibinfo {author} {\bibfnamefont {A.}~\bibnamefont
			{Niederle}}\ and\ \bibinfo {author} {\bibfnamefont {H.}~\bibnamefont
			{Rieger}},\ }\bibfield  {title} {\bibinfo {title} {Superfluid clusters,
			percolation and phase transitions in the disordered, two-dimensional
			{Bose}-{Hubbard} model},\ }\href@noop {} {\bibfield  {journal} {\bibinfo
			{journal} {New J. Phys.}\ }\textbf {\bibinfo {volume} {15}},\ \bibinfo
		{pages} {075029} (\bibinfo {year} {2013})}\BibitemShut {NoStop}%
	\bibitem [{\citenamefont {Schir{\'o}}\ and\ \citenamefont
		{Fabrizio}(2010)}]{Schir0_2010_prl}%
	\BibitemOpen
	\bibfield  {author} {\bibinfo {author} {\bibfnamefont {M.}~\bibnamefont
			{Schir{\'o}}}\ and\ \bibinfo {author} {\bibfnamefont {M.}~\bibnamefont
			{Fabrizio}},\ }\bibfield  {title} {\bibinfo {title} {Time-dependent mean
			field theory for quench dynamics in correlated electron systems},\
	}\href@noop {} {\bibfield  {journal} {\bibinfo  {journal} {Phys. Rev. Lett.}\
		}\textbf {\bibinfo {volume} {105}},\ \bibinfo {pages} {076401} (\bibinfo
		{year} {2010})}\BibitemShut {NoStop}%
	\bibitem [{\citenamefont {Bloch}\ \emph {et~al.}(2008)\citenamefont {Bloch},
		\citenamefont {Dalibard},\ and\ \citenamefont {Zwerger}}]{Bloch_2008_rmp}%
	\BibitemOpen
	\bibfield  {author} {\bibinfo {author} {\bibfnamefont {I.}~\bibnamefont
			{Bloch}}, \bibinfo {author} {\bibfnamefont {J.}~\bibnamefont {Dalibard}},\
		and\ \bibinfo {author} {\bibfnamefont {W.}~\bibnamefont {Zwerger}},\
	}\bibfield  {title} {\bibinfo {title} {Many-body physics with ultracold
			gases},\ }\href@noop {} {\bibfield  {journal} {\bibinfo  {journal} {Rev. Mod.
				Phys.}\ }\textbf {\bibinfo {volume} {80}},\ \bibinfo {pages} {885} (\bibinfo
		{year} {2008})}\BibitemShut {NoStop}%
	\bibitem [{\citenamefont {Zakrzewski}(2005)}]{Zakrzewski_2005_pra}%
	\BibitemOpen
	\bibfield  {author} {\bibinfo {author} {\bibfnamefont {J.}~\bibnamefont
			{Zakrzewski}},\ }\bibfield  {title} {\bibinfo {title} {Mean-field dynamics of
			the superfluid-insulator phase transition in a gas of ultracold atoms},\
	}\href@noop {} {\bibfield  {journal} {\bibinfo  {journal} {Phys. Rev. A}\
		}\textbf {\bibinfo {volume} {71}},\ \bibinfo {pages} {043601} (\bibinfo
		{year} {2005})}\BibitemShut {NoStop}%
	\bibitem [{dat()}]{data}%
	\BibitemOpen
	\href@noop {} {}\bibinfo {note} {Reentrant Bose glass and quench dynamics
		Data,
		\url{https://github.com/ShihaoDing2001/Reentrant_Bose_glass_and_quench_dynamics_Data}.}\BibitemShut
	{Stop}%
	\bibitem [{\citenamefont {Barman}\ \emph {et~al.}(2013)\citenamefont {Barman},
		\citenamefont {Dutta}, \citenamefont {Khan},\ and\ \citenamefont
		{Basu}}]{Barman_2013_epjb}%
	\BibitemOpen
	\bibfield  {author} {\bibinfo {author} {\bibfnamefont {A.}~\bibnamefont
			{Barman}}, \bibinfo {author} {\bibfnamefont {S.}~\bibnamefont {Dutta}},
		\bibinfo {author} {\bibfnamefont {A.}~\bibnamefont {Khan}},\ and\ \bibinfo
		{author} {\bibfnamefont {S.}~\bibnamefont {Basu}},\ }\bibfield  {title}
	{\bibinfo {title} {Understanding the {Bose} glass phase via a percolation
			scenario},\ }\href@noop {} {\bibfield  {journal} {\bibinfo  {journal} {Eur.
				Phys. J. B}\ }\textbf {\bibinfo {volume} {86}},\ \bibinfo {pages} {308}
		(\bibinfo {year} {2013})}\BibitemShut {NoStop}%
\end{thebibliography}
\end{document}